\pdfoutput=1

\documentclass[sigconf,screen,table]{acmart}
\usepackage{amsmath}
\usepackage{graphicx}
\usepackage{booktabs}
\usepackage{latexsym}	% \Diamond
\usepackage{array}		% p{}
\usepackage{multirow}
\usepackage{subcaption}
\usepackage{algorithm}
\usepackage[noend]{algpseudocode}
\usepackage{threeparttable}
\usepackage{paralist}   % compactitem
\usepackage{xspace}
\usepackage{color}
\usepackage{adjustbox}
\usepackage{balance}
\usepackage{wasysym}
\usepackage{rotating}   % turn
\usepackage{listings}
\usepackage{tabu}
\usepackage{pifont}
\usepackage{framed}
\usepackage[shortlabels]{enumitem}

\usepackage{url}            % new ACM
\usepackage{hyperref}

\usepackage{microtype}

\usepackage{xpatch}
\makeatletter
\chardef\TPT@@@asteriskcatcode=\catcode`*
\catcode`*=11
\xpatchcmd{\threeparttable}
  {\TPT@hookin{tabular}}
  {\TPT@hookin{tabular}\TPT@hookin{tabu}}
  {}{}
\catcode`*=\TPT@@@asteriskcatcode
\makeatother

\setcopyright{acmcopyright}
\acmPrice{15.00}
\acmDOI{10.1145/3540250.3549105}
\acmYear{2022}
\copyrightyear{2022}
\acmSubmissionID{fse22main-p227-p}
\acmISBN{978-1-4503-9413-0/22/11}
\acmConference[ESEC/FSE '22]{Proceedings of the 30th ACM Joint European Software Engineering Conference and Symposium on the Foundations of Software Engineering}{November 14--18, 2022}{Singapore, Singapore}
\acmBooktitle{Proceedings of the 30th ACM Joint European Software Engineering Conference and Symposium on the Foundations of Software Engineering (ESEC/FSE '22), November 14--18, 2022, Singapore, Singapore}

\newcommand{\green}[1]{\textcolor[rgb]{0.00,0.60,0.00}{#1}}
\newcommand{\darkblue}[1]{\textcolor[rgb]{0.00,0.00,0.65}{#1}}

\newcommand{\gou}{\green{\ding{52}}\xspace}
\newcommand{\ling}{\darkblue{\RIGHTcircle}\xspace}

\definecolor{wheat1}{rgb}{1.000000,0.905882,0.729412}
\definecolor{LightGray}{rgb}{0.827451,0.827451,0.827451}

\newcolumntype{a}{>{\columncolor{wheat1}}l}

\definecolor{mygreen}{rgb}{0,0.6,0}
\definecolor{mygray}{rgb}{0.5,0.5,0.5}
\definecolor{mymauve}{rgb}{0.58,0,0.82}
\definecolor{darkblue}{rgb}{0.0,0.0,0.6}
\definecolor{maroon}{RGB}{102, 0, 0}
\definecolor{Maroon}{cmyk}{0,0.87,0.68,0.32}
\definecolor{darkred}{RGB}{139, 0, 0}
\definecolor{forestgreen}{RGB}{34, 139, 34}

\lstdefinelanguage{XML}
{
  basicstyle=\ttfamily\small,   %\small is less than \footnote?
  morestring=[b]",
  moredelim=[s][\color{darkblue}]{<}{\ },
  moredelim=[s][\color{darkblue}]{</}{>},
  moredelim=[l][\color{darkblue}]{/>},
  moredelim=[l][\color{darkblue}]{>},
  morecomment=[s]{<?}{?>},
  morecomment=[s]{<!--}{-->},
  stringstyle=\color{darkred},
  identifierstyle=\color{mymauve}
}

\lstdefinestyle{customJava}{
  breaklines=true,
  keepspaces=true,
  frame=single,
  language=Java,
  showstringspaces=false,
  basicstyle=\footnotesize\ttfamily,
  keywordstyle=\color{blue},
  otherkeywords={+, getIntent},
  numbers=left,
  numbersep=5pt,
  numberstyle=\scriptsize\color{black},
  rulecolor=\color{black},
  stepnumber=1,
  tabsize=2,
  commentstyle=\itshape\color{green!40!black},
  stringstyle=\color{orange},
  emph=[1]  %http://tex.stackexchange.com/a/148194
  {
        do,
        try,
        new,
        catch,
        while,
        SecProvider,
        SecReceiver,
        SecService,
        SecActivity,
        SecSink,
  },
  emphstyle=[1]{\color{darkred}},
  emph=[2]  %http://tex.stackexchange.com/a/148194
  {
        @Override,
  },
  emphstyle=[2]{\color{purple!40!black}},
  belowskip=-1em, %http://tex.stackexchange.com/a/50108
}

\newif\ifANNOYMIZE
\ANNOYMIZEfalse

\newif\ifACM
\ACMtrue  %ACM
\ifACM
\fi

\ifACM
\newcommand{\myfig}{Figure\xspace}
\else
\newcommand{\myfig}{Fig.\xspace}
\fi

\ifACM
\newcommand{\mysec}{\S}
\else
\newcommand{\mysec}{Section\xspace}
\fi

\newsavebox{\bigimage} % for positioning subfigures

\begin{document}

% Title
\title{Understanding Blockchain Systems' Vulnerabilities through Clustering, Correlation, and 0-day Discovery}
\title[An Empirical Study of Blockchain System Vulnerabilities]{Diving Into the Weaknesses of Blockchain: An Empirical Study of Blockchain System Vulnerabilities}
\title[An Empirical Study of Blockchain System Vulnerabilities]{An Empirical Study of Blockchain System Vulnerabilities: Modules, Types, and Patterns}
%\subtitle{This is the extended technical report version of our SCLib paper in ACM CODASPY 2018~\cite{SCLib18}.}

\author{Xiao Yi}
\affiliation{\institution{The Chinese University of Hong Kong}\city{Hong Kong SAR}\country{China}}
\email{yx019@ie.cuhk.edu.hk}

\author{Daoyuan Wu}
\authornote{Corresponding author.}
\affiliation{\institution{The Chinese University of Hong Kong}\city{Hong Kong SAR}\country{China}}
\email{dywu@ie.cuhk.edu.hk}

\author{Lingxiao Jiang}
\affiliation{\institution{Singapore Management University}\city{Singapore}\country{Singapore}}
\email{lxjiang@smu.edu.sg}

\author{Yuzhou Fang}
\affiliation{\institution{The Chinese University of Hong Kong}\city{Hong Kong SAR}\country{China}}
\email{yzfang@ie.cuhk.edu.hk}

\author{Kehuan Zhang}
\affiliation{\institution{The Chinese University of Hong Kong}\city{Hong Kong SAR}\country{China}}
\email{khzhang@ie.cuhk.edu.hk}

\author{Wei Zhang}
\affiliation{\institution{Nanjing University of Posts and Telecommunications}\city{Nanjing}\country{China}}
\email{zhangw@njupt.edu.cn}

% The default list of authors is too long for headers}
% \renewcommand{\shortauthors}{X. Yi et al.}

% !TeX root = main.tex

\begin{abstract}

Blockchain, as a distributed ledger technology, becomes increasingly popular, especially for enabling valuable cryptocurrencies and smart contracts.
%an emerging technology for its decentralization
%and the capability of enabling cryptocurrencies and smart contracts.
%However, as a distributed ledger software by nature, blockchain inevitably has software issues.
%While recent works investigated the security of application-level smart contracts, the underlying system-level security bugs of blockchain have \textit{not} been systematically studied. 
However, the blockchain software systems inevitably have many bugs.
Although bugs in smart contracts have been extensively investigated,
security bugs of the underlying blockchain systems are much less explored. 
In this paper, we conduct an empirical study on blockchain's system vulnerabilities from four representative blockchains, Bitcoin, Ethereum, Monero, and Stellar.
%As the first major contribution, 
Specifically,
%Due to the lack of CVE information associated with these blockchain projects, 
we first design a systematic filtering process to effectively identify 1,037 vulnerabilities and their 2,317 patches from 34,245 issues/PRs (pull requests) and 85,164 commits on GitHub.
We thus build the first blockchain vulnerability dataset, which is available at \url{https://github.com/VPRLab/BlkVulnDataset}.
%Our second major contribution is 
We then perform
    unique analyses of this dataset
    %across different blockchain projects
    at three levels, including
(i) file-level vulnerable module categorization by identifying and correlating module paths across projects, 
(ii) text-level vulnerability type clustering by natural language processing and similarity-based sentence clustering, 
and (iii) code-level vulnerability pattern analysis by generating and clustering code change signatures that capture both syntactic and semantic information of patch code fragments.
%text-level vulnerability type clustering by using natural language processing to recognize type keywords and then clustering them

%Besides pinpointing six susceptible blockchain modules and identifying 20 blockchain vulnerability types (including seven non-traditional ones) that affect at least ten vulnerabilities each, our major finding is the discovery of 21 blockchain-specific vulnerability patterns, which are obtained from the 174 clusters of code change signatures generated from 3,251 code fragments.
%These patterns check unique blockchain attributes and validate various blockchain statuses during node synchronization, peer validation, wallet and\linebreak database operations,
%and can guide vulnerability detection/fixing.
%%\lxedit{which can be useful for detecting and fixing new vulnerabilities of the same types and patterns.}
%%We also discuss and demonstrate their usages.

%Among detailed results, 
Our analyses reveal three key findings:
(i) some blockchain modules are more susceptible than the others; notably, each of the modules related to consensus, wallet, and networking has over 200 issues;
%\lxnote{in comparison, what's the number for other modules?}
%than other modules, likely due to their higher code complexity;
(ii) about 70\% of blockchain vulnerabilities are of traditional types, but we also identify four new types specific to blockchains;
and (iii) we obtain 21 blockchain-specific vulnerability patterns that capture unique blockchain attributes and statuses,
%and leverage them to discover 23 similar vulnerabilities in other top blockchains, such as Dogecoin and Bitcoin SV; most of them have been confirmed by their vendors.
and demonstrate that they can be used to detect similar vulnerabilities in other popular blockchains, such as Dogecoin, Bitcoin SV, and Zcash.

\end{abstract}




\maketitle

%\lxnote{it may be tricky to claim 'First' empirical study in the title/abstract; may be better to say 'In-Depth'. The 'first' blockchain vulnerability dataset may be changed to the 'largest'...to date.}

% !TeX root = main.tex

% \vspace{-1ex}
\section{Introduction}
\label{sec:intro}
% \vspace{-0.5ex}

While blockchain was first invented as a transaction ledger of the Bitcoin cryptocurrency~\cite{nakamoto2019bitcoin}, it is now serving as a fundamental component of many cryptocurrencies, 
the total market capitalization of which is close to two trillion USD in February 2022~\cite{cryptocap2022}.
%early April 2021~\cite{cryptocap}. %https://www.bloomberg.com/news/articles/2021-04-05/crypto-market-cap-doubles-past-2-trillion-after-two-month-surge
% Hyperledger fabric: a distributed operating system for permissioned blockchains.
Smart contract platforms (e.g., Ethereum~\cite{buterin2014next} and Hyperledger Fabric~\cite{androulaki2018hyperledger}) and decentralized computing platforms (e.g., Interplanetary File System~\cite{benet2014ipfs} and Blockstack~\cite{ali2016blockstack}) %Juan Benet. 2014. IPFS - Content Addressed, Versioned, P2P File System. CoRR arXiv abs/1407.3561; Blockstack: A Global Naming and Storage System Secured by Blockchains
further evolved the blockchain technology into various decentralized applications,
such as DeFi (Decentralized Finance)~\cite{werner2021sok}, %https://arxiv.org/abs/2101.08778
smart contract oracles~\cite{zhang2020deco, zhang2016town}, %DECO (CCS'20), Town Crier (CCS'16)
decentralized identities~\cite{maram2020candid}, % https://www.fanzhang.me/publications/candid/
decentralized IoT management~\cite{IoTPassport19}, %~\cite{IoTBlockchain20} %~\cite{zhang2018smart}, 
and decentralized app markets~\cite{chen2021agchain}. % https://arxiv.org/abs/2101.06454
To protect the decentralization of these systems and secure those finance-critical cryptocurrencies, security is a top priority of many blockchains.

Prior research on blockchain security focused on smart contract vulnerability detection and network analysis.
Many static program analysis tools, e.g., Oyente~\cite{luu2016smarter}, Zeus~\cite{Kalra2018ZEUSAS}, Securify~\cite{tsankov2018securify}, Gigahorse~\cite{grech2019gigahorse}, and ETHBMC~\cite{ETHBMC20}, have been proposed to detect vulnerable smart contracts via symbolic execution and model checking.
Dynamic tools~\cite{jiang2018contractfuzz, Sereum19, SODA21, ConFuzzius21} and learning-based tools \cite{SmartEmbed20, ESCORT21, AMEVulDetector21} were also invented. % such as ContractFuzzer~\cite{jiang2018contractfuzz} and Sereum~\cite{Sereum19}.
Besides smart contract analysis, some works analyzed network traffic hijacking~\cite{apostolaki2017hijackbitcoin} and mining~\cite{gao2019minebitcoin} attacks and performed transaction attack analysis~\cite{karme2012fastbitcoin, EthereumGraphAnalysis18, Zhang2020TXSPECTORUA, DeFiPoser21}.
In contrast, blockchains' system-level security issues are much less explored in academic research. 
To the best of our knowledge, there was only one study~\cite{wan2017blkbugs}
% (from the software engineering community) 
in this direction.
It specifically analyzed 946 blockchain bugs, with only 18 security bugs covered and four analyzed.

In this paper, we aim to \textit{systematically understand blockchain system vulnerabilities} by conducting an empirical vulnerability study of the representative blockchains in four directions, including the classic Bitcoin~\cite{nakamoto2019bitcoin}, the smart contract platform Ethereum~\cite{buterin2014next}, the anonymous coin Monero~\cite{Noether2015RingSC}, and the payment network Stellar~\cite{lokhava2019stellar}.
They are not only popular in the cryptocurrency market but also backed up with solid technical papers.
%\myfig\ref{fig:overview} shows a high-level overview of our study, consisting of three major parts: data collection, vulnerability analysis, and result presentation.
%We now explain them as follows.

\begin{figure*}[t!]
% \vspace{-2ex}
\begin{adjustbox}{center}
\includegraphics[width=1.0\linewidth]{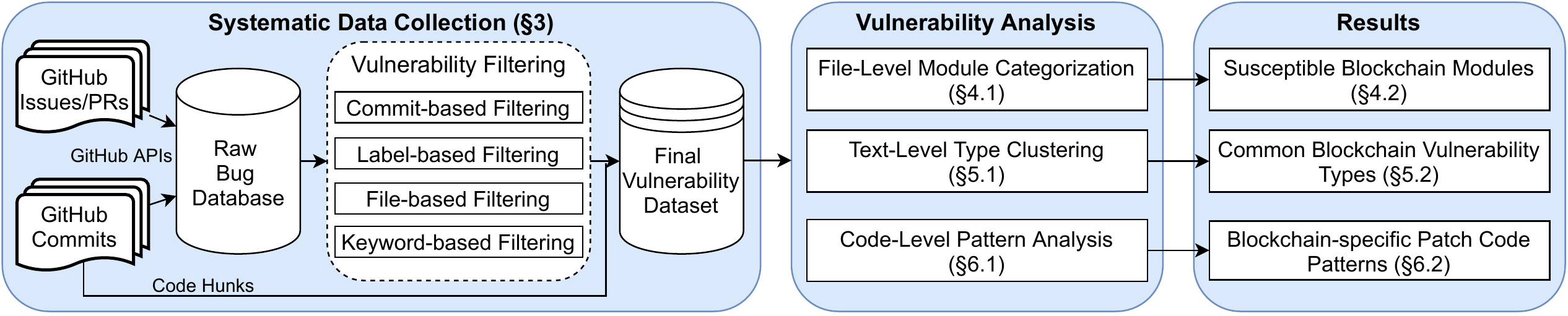}
\end{adjustbox}
% \vspace{-4ex}
\caption{The overall workflow of our blockchain vulnerability study.}
% \vspace{-2ex}
\label{fig:overview}
\end{figure*}

As depicted in \myfig\ref{fig:overview}, the first step and challenge of our study is to \textit{effectively} collect vulnerable issues and their patches of those four blockchains.
This is difficult because there is very little CVE information associated with blockchain projects (unlike other vulnerability mining studies~\cite{PatchStudy17, WooyunHackerOne15, AndroVulns19}), and the large number (over 34K) of raw blockchain bugs in our crawled database makes manual vulnerability filtering\footnote{That said, we need to recognize or differentiate \textit{real} vulnerabilities from \textit{regular} bugs.} ineffective.
To address this, we propose a vulnerability filtering framework based on the intuition that vulnerabilities have unique characteristics from various aspects, and we can gradually identify candidate vulnerabilities by analyzing attributes of the code commits, files, labels, and keywords.
%from coarse-grained to fine-grained levels.
%\lxnote{avoid overloading the ``level'' word in different contexts; only use it for file-/text-/code-levels}
%Specifically, we perform the filtering at the commit, file, label, and keyword levels.
Eventually, we obtain 1,037 vulnerabilities and their 2,317 patches as our blockchain vulnerability dataset.

Based on this unique dataset, we study three key yet unexplored aspects of blockchain vulnerabilities, including susceptible blockchain modules, common blockchain vulnerability types, and blockchain-specific patch code patterns.
To this end, we perform the file-, text-, and code-level vulnerability analysis as follows.

%\lxnote{the following can be better in making the analysis results more prominent than the analysis methods, like the abstract: may describe the 3 levels of analysis methods first briefly; after that, present the analysis results.} 
\textit{Firstly}, we conduct the module analysis by inspecting patched files.
However, inspecting each individual file is time-consuming because there are 2,362 unique patch file paths.
Therefore, we propose to identify the \textit{module path}, i.e., the folder name that could summarize the module of enclosed files (e.g., the ``rpc/'' folder indicates the RPC module).
We further correlate module paths across different blockchains by identifying a reference blockchain architecture and mapping different module paths into this architecture.
This module categorization allows us to obtain a layered map of blockchain vulnerabilities in different modules and pinpoint susceptible blockchain modules.
We find that some modules are more susceptible than the others, such as the highly susceptible ones related to consensus, wallet, and networking, each with over 200 vulnerabilities.
%Other modules on RPC, GUI/CMD, and storage are also susceptible.
%\TODO{\red{Remove -}}
%Besides they are commonly used in blockchain systems, we further show that the higher patch code complexity could be another factor causing them to be more susceptible.
%\blue{Moreover, we further analyze and compare the vulnerability module differences across different blockchains.}

\textit{Secondly}, we perform the type analysis by analyzing vulnerability text, more specifically, vulnerability titles.
This is because a vulnerability type is typically captured by the title of an issue/PR (pull request), e.g., Bitcoin PR \#17640 ``wallet: Fix uninitialized read in bumpfee(…),'' where ``uninitialized read'' is the type.
To eliminate noisy words and generate good-quality clusters about types, we leverage the part-of-speech analysis of NLP (natural language processing) to first extract \textit{type keywords} before we conduct actual clustering.
%, based on the grammatical pattern that a type is often located in between a verb (e.g., ``fix'') and a preposition (e.g., ``in'').
By extracting type keywords in various situations and identifying a suitable clustering algorithm (and its setting), we successfully map 75.8\% of the vulnerabilities into the clusters of different types and analyze the top 20 types that affect at least ten vulnerabilities each.
Among these types, we identify four new vulnerability types that are directly related to blockchain transaction, block, peer/node, and wallet key/password.
%four are specific to blockchains and three are partially specific, whereas traditional vulnerability types still hold 62\%$\sim$78\% of all the vulnerabilities. 
We also show that traditional vulnerability types still hold 62\%$\sim$78\% of all the blockchain vulnerabilities.
Furthermore, we analyze the type differences across different blockchain projects.

\textit{Thirdly}, we conduct the pattern analysis by analyzing vulnerability patch code.
In particular, we focus on blockchain-specific vulnerability types since the code patterns of traditional vulnerability types are well-known.
To facilitate similar patch code into the same cluster, we design and generate the \textit{code change signatures} that concisely capture both syntactic and semantic information of patch code fragments.
By clustering 3,251 code fragments into 174 clusters of code change signatures, we identify 21 blockchain-specific vulnerability patterns that check unique blockchain attributes (e.g., the sender address, transaction order, block header, and gas limit) and validate various blockchain statuses during node synchronization, peer validation, wallet, and database operations.
%These patterns are generalized across four blockchain systems in a few different programming languages (e.g., C/C++ and GO)
%and can be used to detect and fix new vulnerabilities of the same types and patterns.
We further leverage these patterns to discover 20 similar vulnerabilities in other popular blockchains, notably, Dogecoin, Bitcoin SV, and Zcash, which have a collective market capitalization of over 25 billion USD as of January 2022.
Most of our vulnerability reports have been confirmed and are under patching, with only two being rejected.
This demonstrates the real-world impact of our vulnerability patterns.
A thorough detection of blockchain system vulnerabilities based on the patterns extracted in this paper will be our future work.

To sum up, the main contributions of this paper are as follows:
\begin{itemize}[nosep,leftmargin=1em]
\item We design a systematic filtering process to curate a unique vulnerability dataset and will release it to the research community.
The link of the dataset is already available at \url{https://github.com/VPRLab/BlkVulnDataset}.
% and relevant analysis scripts.

\item We develop a set of new methods to analyze blockchain vulnerabilities, build a knowledge base on previously unknown patterns of the vulnerabilities and their fixes.

\item We reveal three key findings about blockchain system vulnerabilities in terms of their susceptible modules, various vulnerability types, and specific vulnerability patterns.
Moreover, we demonstrate the usage of these vulnerability patterns by detecting 20 similar vulnerabilities in other popular blockchains. %in Dogecoin, Bitcoin SV, and Zcash.

\end{itemize}

%\lxnote{Overall it may be better to still include the `paper organization' paragraph, especially to make it clearer how Section 4,5,6 are related.} 
\noindent
The rest of this paper is organized as follows.
We first provide the background of studied blockchains and their bug-fixing process in \mysec\ref{sec:backg} and describe our systematic data collection in \mysec\ref{sec:collect}.
We then present our multi-level vulnerability analysis in \mysec\ref{sec:categorize}, \mysec\ref{sec:cluster}, and \mysec\ref{sec:detect}, respectively.
\mysec\ref{sec:related} summarizes the related works.
Finally, \mysec\ref{sec:conclusion} concludes this study.

% !TeX root = main.tex

% \vspace{-1.5ex}
\section{Background}
\label{sec:backg}
% \vspace{-0.5ex}

%In this section, we introduce the background of four representative blockchains we study in this paper and the typical bug-fixing process in these blockchain projects.

% TODO
%\begin{figure*}[t!]
%\vspace{-2ex}
%\begin{adjustbox}{center}
%\includegraphics[width=0.95\linewidth]{issue_commit_relation.pdf}
%\end{adjustbox}
%\vspace{-4ex}
%\caption{The relationship between three GitHub concepts: commit, issue, and pull request.}
%\vspace{-2ex}
%\label{fig:issue_exp}
%\end{figure*}

%\vspace{-1ex}
\subsection{Four Representative Blockchains Studied}
\label{sec:backgblockchain}
%\vspace{-1ex}
%\lxnote{Updated to as of the end of April. If space is limited, much info in the following paragraphs can be shortened.}

In this paper, we study the representative blockchains that are (i) popular in the cryptocurrency market, (ii) in different directions of blockchain usages, and (iii) backed up with solid technical papers.
%\lxnote{hard to say the 3 conditions are really match...}
Under these three conditions, we select the classic Bitcoin~\cite{nakamoto2019bitcoin}, the smart contract platform Ethereum~\cite{buterin2014next}, the anonymous coin Monero \cite{Noether2015RingSC}, and the payment network Stellar~\cite{lokhava2019stellar}.
Next, we present their basic information and the development status on GitHub.

\textbf{Bitcoin}
%is the first decentralized cryptocurrency and often described as ``digital gold.''
%Bitcoin
introduces the concept of \textit{blockchain} \cite{nakamoto2019bitcoin} and uses it as a distributed ledger to record transactions for public verification.
As of January 2022, the Bitcoin cryptocurrency (or BTC) has the top one market capitalization of more than 832 billion USD.
The Bitcoin software was released in 2009, and it is actively maintained by over 850 contributors on GitHub in a repository called \textbf{bitcoin/bitcoin}.
The primary programming language of Bitcoin is C++.

\textbf{Ethereum} is the first blockchain system with the capability of constructing Turing-complete \textit{smart contracts}~\cite{buterin2014next}, which contain a set of pre-defined rules and regulations for self-execution.
To maintain the operation of Ethereum, it creates a native cryptocurrency called Ether (or ETH), which is the second largest cryptocurrency with a market capitalization of more than 410 billion USD as of January 2022.
The Ethereum software was released on GitHub in 2015, and its Go implementation is maintained by over 700 contributors in a repository called \textbf{ethereum/go-ethereum}.

\textbf{Monero} aims to mitigate the privacy leakage in blockchain systems, since each blockchain transaction is transparent and could leak some sensitive information. 
To do so, Monero uses an obfuscated ledger~\cite{Noether2015RingSC} to prevent the transaction details (e.g., transaction source, amount, and destination) from being revealed to outside observers. 
As of January 2022, the Monero coin (XMR) is ranked 47th with a market capitalization of over 3.8 billion USD.
The Monero software was released on GitHub in 2014, and it is maintained by over 250 contributors in a repository called \textbf{monero-project/monero}.
The primary language of Monero is C++.

\textbf{Stellar} is a blockchain-based \textit{payment network}~\cite{lokhava2019stellar} that can perform cross-border money transfer in seconds.
It uses a novel consensus protocol called \textit{Stellar Consensus Protocol} (SCP)~\cite{lokhava2019stellar} for fast and secure transactions among untrusted participants.
The native cryptocurrency of Stellar is called XLM, which is ranked 30th with a market capitalization of around 6.7 billion USD as of January 2022.
The Stellar software was released on GitHub in 2015, and it is currently maintained by more than 80 contributors in a repository called \textbf{stellar/stellar-core}.
Similar to Bitcoin and Monero, the primary language of Stellar is also C++.

\subsection{Bug-fixing Process in Blockchain Projects}
\label{sec:backgbugfixing}
% \vspace{-0.5ex}

It is also necessary to understand the typical bug-fixing process of blockchain projects hosted as open-source projects on GitHub
in order to collect and analyze their vulnerabilities and patches. 
%There are three concepts, commits, issues, and pull requests. 
A \textbf{commit} is a set of changes submitted by developers into a project repository;
a commit can change anything, ranging from changing source code to modifying document files or merging multiple previous commits.
A change consisting of a consecutive sequence of added/deleted lines is also known as a \textbf{hunk}.
A \textbf{patch} is a collection of changes or commits that can be applied to a set of files via a patching tool.
An \textbf{issue} is often a report on a project's GitHub page; it may describe a potential bug or sometimes an enhancement or a question, and may come with fixes and solutions. 
A \textbf{pull request (PR)} is the proposed commit for a project from a separate clone of the project;
it can be pulled from the project clone and accepted into the original project based on the review of managing developers.
For simplicity, we do not explicitly distinguish an issue and a PR in this paper since the latter often contains a bug description too.
Indeed, GitHub itself mixes up the usage of issue/PR numbers.

\section{Systematic Data Collection}
\label{sec:collect}
% \vspace{-0.5ex}

As shown in \myfig~\ref{fig:overview}, the first and a critical step of our study is to collect a good-quality blockchain vulnerability dataset across multiple blockchain systems
that satisfies two conditions:
(i) cover as \textit{many} vulnerabilities as possible in the studied blockchains (i.e., minimizing false negatives);
and (ii) introduce as \textit{few} non-vulnerability bugs as possible in the dataset (i.e., minimizing false positives).

%There can be various ways to collect vulnerability data from project outsiders' perspectives.
Some other vulnerability studies~\cite{PatchStudy17, WooyunHackerOne15, AndroVulns19}
leverage the CVE (Common Vulnerabilities and Exposures) or Bulletin (i.e., bug bounty) information to collect vulnerability data. 
However, we found that there is very little CVE/Bulletin information about most blockchains because blockchain vulnerabilities are critical and often patched directly via the reports from bug bounty programs without releasing a CVE.
For example, Ethereum (go-ethereum) had only four CVEs released before our data collection while Bitcoin had 33 CVEs.
%, which are significantly fewer than what we collect in this paper. 

We take a different way---directly analyze the blockchain projects' issues and commits in their GitHub repositories and extract the vulnerable ones from them.
%We choose to analyze those blockchains' GitHub repositories (i.e., the repositories highlighted in~\mysec\ref{sec:backgblockchain}) since they often provide detailed bug and vulnerability descriptions in their issues or PRs and patch code in the commits related to the issues or PRs (see~\mysec\ref{sec:backgbugfixing}).
%We investigated many data sources, including bounty programs, the Common Vulnerabilities and Exposures (CVE)~\cite{cve} website, and the National Vulnerability Database (NVD)~\cite{nvd}. 
%However, they are not well-maintained for blockchain vulnerabilities, and the provided information is insufficient for our analysis. 
%For example, bounty programs such as HackerOne~\cite{hackerone} and BugCrowd~\cite{bugcrowd} only disclose reported vulnerabilities after a specific time when the issue is fixed for security consideration.
%CVE and NVD are not regularly updated for blockchain system vulnerabilities and only contain simple descriptive messages without patch codes.
%Thus, we choose Github repositories as our data sources, since they are actively maintained and provide detailed vulnerability descriptions (i.e., issues/PRs) and patch codes (i.e., commits).
%After collecting all issues/PRs and commits, another challenge we are facing here is how we can accurately identify vulnerability-related fixes among all the collected issues/PRs and their commits.
%To this end, 
We first crawl all blockchain bugs and organize them into a raw bug database (in \mysec\ref{subsec:crawl}).
The major challenge is how to recognize or differentiate \textit{real} vulnerabilities\footnote{In this paper, we adopt a broad definition of vulnerabilities that considers \textit{the bugs with security impact} as vulnerabilities.} from a large number of \textit{regular} bugs.
%TODO more refs? 2103.13375
To address the challenge, we propose a novel vulnerability filtering framework (in \mysec\ref{subsec:filter}) that systematically and effectively filters out regular bugs and extracts blockchain vulnerabilities.
%By applying this novel method to the raw database, 
We eventually obtain the first
dataset of blockchain system vulnerabilities (in \mysec\ref{subsec:dataset}), comprising more than 1K vulnerabilities identified from over 34K issues.
%from four major blockchains.
It could not be done via manual analysis or via prior training-based patch identification~\cite{IdentifyPatch12, AutoFilterVuln17} since 
(i) there is no ground-truth training set for blockchain vulnerabilities and 
(ii) the learning-based nature of those techniques tends to identify only the similar bugs or vulnerabilities.
%Moreover, the large number of blockchain bugs in our raw database, over 34K, makes manual analysis ineffective.

% \vspace{-1.5ex}
%\subsection{Crawling and Organizing a Comprehensive Blockchain Bug Database}
\subsection{Crawling and Organizing Blockchain Bugs}
\label{subsec:crawl}
% \vspace{-0.5ex}

As illustrated in \myfig~\ref{fig:overview}, our blockchain bug database is constructed from two data sources, the \textit{issues} and \textit{commits},
by leveraging GitHub APIs\footnote{\url{https://docs.github.com/en/rest/reference/commits} and \url{https://docs.github.com/en/rest/reference/issues}}.
For the issues, we collect all the information of each closed issue/PR, including the issue title, issue body, comments, events, and bug category labels.
We consider only \textit{closed} issues/PRs because \textit{open} issues are not confirmed bugs yet and certainly have no patches.
Note that even for closed issues, they may not be the real bugs and could have no patches (i.e., they were simply closed by developers). 
For the commits, we first crawl
%\footnote{Since the metadata of \textit{issues} and the information for correlating a commit and an issue are available only on GitHub, we use crawling to handle them, as well as \textit{commits}.} 
\textit{all} the commits of a repository and then determine which commits are bug-related.
%we use the same crawling method to handle \textit{commits} instead of interacting with the cloned git repositories.
For each commit, we collect its title, commit message, affected files, and id/URL; for some commits filtered according to \mysec\ref{subsec:filter}, their actual code change hunks are also collected and processed according to \mysec\ref{subsec:dataset} and used for code-level pattern analysis in \mysec\ref{sec:detect}.
%For these raw data, we can leverage GitHub APIs to crawl them directly without composing a web page crawler.
%Based on the crawling principle described here, 
We have collected a total of 34,245 \textit{closed} issues/PRs and 85,164 commits as the raw dataset at the end of February 2020.
The detailed breakdown of these issues/PRs and commits across four blockchain projects is available in Table~\ref{tab:repo_info}.

With the raw data collected, a non-trivial task is to organize and correlate the issues with their corresponding commits.
Specifically, we need to determine all the relevant commits for a given issue/PR --- if an issue/PR has no patch commits, it is not a real bug and will be filtered out.
By summarizing the issue/PR and commit's GitHub structures, we observe three kinds of information we can leverage for such correlation.
First, we leverage the issue page's event information (e.g., XXX mentioned this issue and YYY added a commit) and retrieve the commit URLs from those events.
For example, in \url{https://github.com/bitcoin/bitcoin/issues/595}, we obtain the commit URL via the event of ``laanwj added a commit that referenced this issue.''
Second, for a PR like \url{https://github.com/bitcoin/bitcoin/pull/9366}, we can directly retrieve its commit lists at its ``Commits'' tab page.
Although these two kinds of information is useful for most issues/PRs, some commits may not appear in the events of issues or commit lists of PRs.
To overcome this, our script analyzes all the 85,164 commits' titles and messages and identifies issue/PR numbers from them.
%TODO \lxnote{unclear to me why we need to craw the commit web pages, instead of retrieving from the git repo? May simplify the description of how the information is retrieved/correlated; just describe what information is needed for the correlation.}
%TODO \red{(Do we give an example that the issue page does not mention the commits but we can identify the issue from the commit page?)}
% https://github.com/bitcoin/bitcoin/commit/8c6081a884cd0969160955ce8687d4d4ed074db3
% https://github.com/stellar/stellar-core/commit/7d7a409a143ef052d5b3d71b951f32414f8faf36
With these strategies, we successfully build the relationship between the issues and commits and finish constructing the raw bug database shown in \myfig~\ref{fig:overview}.

\begin{comment}
However, the raw bug database did not yet contain patch code,
%\lxnote{this phrase may not be right; isn't that the commits may contain some patch code?}
which is required for code-level pattern analysis in \mysec\ref{sec:detect}.
We are interested in the patch code for the bugs determined as vulnerabilities through our vulnerability filtering in \mysec\ref{subsec:filter}.
Specifically, there are a total of 1,059 vulnerable issues/PRs, which associate with 2,933 code commits.
Our objective is to collect the complete (patch) code hunks of these 2,933 commits, including their added, removed, and neighboring context code lines for future use.
%Unfortunately, GitHub does not provide APIs for fetching commits' code hunks\lxnote{this doesn't sound right; git itself provides such APIs. May just remove this sentence; just say what you've done.}.
%We develop our web scraping tool based on Selenium~\cite{Selenium} to automatically crawl and parse code hunks from commits' HTML pages on GitHub and save them in structured JSON formats for easy reference.
We then develop a script to automatically parse code hunks from patch commits and save them in structured JSON formats for easy reference.
\end{comment}

% \vspace{-2ex}
\subsection{A Vulnerability Filtering Framework}
\label{subsec:filter}
% \vspace{-0.5ex}

To evolve the raw bug database into the final vulnerability dataset, we design a systematic vulnerability filtering framework expressed as a seven-step process (i.e., S0$\sim$S4b in Table~\ref{tab:filter_result}) to effectively differentiate vulnerabilities from regular bugs with minimal manual work.
The intuition is that vulnerabilities have unique characteristics at various aspects, and we can gradually identify candidate vulnerabilities by analyzing attributes of the code commits, files, labels, and keywords.
%bug attributes from coarse-grained to fine-grained levels.
As shown in Table~\ref{tab:filter_result}, we perform the filtering at the following four aspects:
%\lxnote{avoid using ``level'' here to avoid word conflicting with the 3 levels of analysis later.}

\textbf{Commit-based filtering.}
Firstly, in the step S0, we leverage the most straightforward characteristic that \textit{a closed vulnerability must associate with code commits}.
In other words, an issue/PR without any commit could be excluded directly. 
%This kind of issues/PRs are normally general inquiries about problems of project usage, or identified as invalid or duplicated questions by project contributors, or suggestions of project features. 
%Since no commits are related to these issues/PRs, meaning there is no changing of codes, we can be sure that they are unrelated to vulnerability fixes.
Since we have already built the relationship between issues/PRs and commits in \mysec\ref{subsec:crawl}, we easily exclude 10,101 issues/PRs out of the entire 34,245 issues/PRs. %TODO in the raw bug database.
%Our further filtering thus can focus on the remaining 24,144 issues/PRs and their corresponding commits.

\begin{table}[t!]
\caption{Intermediate results of the filtering in each step.}
\label{tab:filter_result}
% \vspace{-3ex}
\begin{adjustbox}{center} %angle=90
\scalebox{0.85}{
	\begin{tabular}{|c|c|c|c|r|l|r|l|}
		\hline
		\multirow{2}{*}{\textbf{Action}} &
		  \textbf{Commit} &
		  \multicolumn{2}{c|}{\textbf{File}} &
		  \multicolumn{2}{c|}{\textbf{Label}} &
		  \multicolumn{2}{c|}{\textbf{Keyword}} \\ \cline{2-8} 
		                & S0  & S1 & S2 & S3a & S3b & S4a & S4b \\ \hline
		\begin{tabular}[c]{@{}c@{}}Include/\\ Exclude\end{tabular} & -10,101         & -3,798      & -1,522      & 56        & -4,400        & 1,227        & -6,330       \\ \hline
		Remain          & 24,144  & 20,346 & 18,824 & 18,768 & 14,368 & 13,141 & 6,811  \\ \hline
	\end{tabular}%
}
\end{adjustbox}
% \vspace{-3ex}
\end{table}

% \begin{itemize}
% 	\item Commit-based filtering: rule 0.
% 	\item File-based filtering: rule 1 and rule 2.
% 	\item Label-based filtering: rule 3 and rule 4.
% 	\item Keyword-based filtering: rule 5 and rule 6.
% \end{itemize}

\textbf{File-based filtering.}
Secondly, we leverage two characteristics of patch files to filter out the bugs that are certainly not vulnerabilities.
The basic idea of these two characteristics is that \textit{the patch of a vulnerable issue/PR must make some real code changes}, including changing files with actual source code and not containing only test code.
Specifically, in the step S1, we determine the file types with actual source code (by their file suffixes) for four blockchains.
An issue/PR whose commits do not modify any file in these types should be excluded.
For example, there are 152 different file types for Bitcoin's commits, but only these seven file types, [`.cpp', `.h', `.py', `.sh', `.cc', `.c', `.java'], contain actual source code whereas other file types like `.yml' and `.mk' are unlikely related to vulnerabilities.
This step filters out 3,798 more issues/PRs, then the remaining 20,346 are further filtered by the step S2.
Specifically, S2 excludes the test-only commits and their associated issues/PRs.
With the file-based filtering, we exclude 22\% (5,322/24,144) of the issues/PRs.
% and can focus on 18,824 issues/PRs in the next steps.\lxnote{can simplify the descriptions.}
%Issues/PRs excluded by rule 1 and rule 2 usually are modifications of documentations, configurations, or supplementary tests, that are not valuable for our analysis.

\textbf{Label-based filtering.}
%Among the remaining 18,824 issues/PRs, we notice that one-third of them are assigned labels to describe their bug natures.
Thirdly, we leverage the characteristic of the labels of issues/PRs: \textit{certain words in the labels could indicate whether an issue/PR is related to a vulnerability or not}.
For example, the `Privacy' label marks privacy-related bugs in the Bitcoin project and the `obsolete:vuln' label indicates the early-stage vulnerabilities of Ethereum.
To avoid false positives, we are conservative in specifying vulnerability labels --- we assign only three labels (i.e., the `Privacy', `obsolete:vuln', and special label `SEC-XXX' that appeared in the beginning of issue/PR titles)
and mark their corresponding 56 issues/PRs explicitly as vulnerabilities in the step S3a.
In contrast, there are much more labels clearly indicating non-vulnerability issues/PRs.
Specifically, out of the entire 87 labels from four blockchain projects, we manually determine that 48 of them are \textit{not} related to vulnerabilities, such as `Refactoring', `Docs', and `type:feature'.
With these labels, we filter out their associated 4,400 issues/PRs in the step S3b.
After this step, we have narrowed the filtering scope from 34,245 to 14,368 issues/PRs, a reduction of 58\%.

\textbf{Keyword-based filtering.}
Lastly, we directly check issues/PRs' text based on the characteristic that \textit{some keywords could indicate an issue/PR vulnerable whereas others could imply an issue/PR not related to vulnerabilities}.
% all_issue_word_count.txt --> all_issue_similar_word.txt --> all_issue_word_record.txt --> key_word_index
To this end, we first perform a word count analysis on the words in issue/PR titles and bodies, sort these words by their appearance frequency, and exclude the words that appear only once.
We then group the words by their semantic similarity using the spaCy~\cite{spacy} NLP library.
Since similar words are grouped together, we manually go through all the clusters to obtain a set of vulnerability-related words (Step S4a) or non-vulnerability words (Step S4b).
Specifically, we obtain 62 clusters of vulnerability-related words and 79 clusters of non-vulnerability words, which allows us to automatically identify 1,227 vulnerable issues/PRs and exclude 6,330 irrelevant issues/PRs in the step S4a and S4b, respectively.

Eventually, our filtering framework extracted 1,283 (=1,227+56) suspicious issues/PRs (in the step S3a and S4a)
from the entire 34,245 issues/PRs.
We have manually examined all these candidates and confirmed that 1,059 of them were actually vulnerability-related.
This suggests that our filtering achieves a precision of 82.5\% in identifying true vulnerabilities.
It is also worth noting that our filtering framework may potentially have a high recall in identifying all patched vulnerabilities in the projects although there is no ground-truth for exact measurement, since it handles at least 80.1\% (27,434/34,245) of all the issues/PRs;
although the remaining 6,811 after step S4b are discarded, we believe that they have a low chance of being vulnerabilities due to no relevant keywords.
%It is worth noting that while our filtering framework targets blockchains only in this paper, it could be applied to other kinds of open-source projects to extract their vulnerability-related issues/PRs.

\begin{table}[t!]
	\caption{Metadata of the raw and vulnerability datasets.}
	\label{tab:repo_info}
	% \vspace{-3ex}
	% \resizebox{\linewidth}{!}{%
	\begin{adjustbox}{center} %angle=90
	\scalebox{0.9}{
		\begin{tabular}{|c|c|c|c|c|} 
			\hline
			\multirow{2}{*}{\textbf{Repository}} &
			  \multicolumn{2}{c|}{\textbf{Raw Bug Database}} &
			  \multicolumn{2}{c|}{\textbf{Vulnerability Dataset}} \\ \cline{2-5} 
			 &
			  \begin{tabular}[c]{@{}c@{}}Closed\\Issues/PRs\end{tabular} &
			  Commits &
			  \begin{tabular}[c]{@{}c@{}}Vulnerable\\Issues/PRs\end{tabular} &
			  \begin{tabular}[c]{@{}c@{}}Patch\\Commits\end{tabular} \\ \hline
			Bitcoin        & 16,731          & 41,706          & 442            & 942            \\
			Ethereum       & 9,321           & 23,764          & 365            & 826            \\
			Monero         & 5,918           & 12,656          & 178            & 286            \\
			Stellar        & 2,275           & 7,038           & 52             & 263            \\ \hline
			\textbf{Total} & \textbf{34,245} & \textbf{85,164} & \textbf{1,037} & \textbf{2,317} \\ \hline
		\end{tabular}
	}
	\end{adjustbox}
% \vspace{-5ex}
\end{table}

% \vspace{-2ex}
\subsection{The Vulnerability Dataset and Its Metadata}
\label{subsec:dataset}
% \vspace{-0.5ex}

%As mentioned in the end of \mysec\ref{subsec:crawl}, 
We then retrieve the actual code hunks for the identified 1,059 issues/PRs from their corresponding 2,933 commits.
This allows us to further exclude 22 issues/PRs because they associate with ``invalid'' code commits through the code hunk analysis.
Specifically, we identified 586 duplicate code commits whose code hunks were the same (e.g., \url{https://github.com/bitcoin/bitcoin/commit/d4781ac6} and \url{https://github.com/bitcoin/bitcoin/commit/8a445c56}), for which we kept just one code commit for each duplicate pair.
% https://github.com/ethereum/go-ethereum/commit/c12180d00566dba977b2d34f61ed52e2a4d279ed
We also found 30 empty code commits where we were not able to obtain their code hunks due to disappeared (e.g., \url{https://github.com/bitcoin/bitcoin/commit/7e193ff6}) or large diffs (e.g., \url{https://github.com/ethereum/go-ethereum/commit/34dde3e2}).  %\lxnote{why large diffs can't be analyzed?}  % HTTP 404
As a result, our final vulnerability dataset consists of 1,037 vulnerability-related issues/PRs and their 2,317 commits, as shown in Table~\ref{tab:repo_info}.
%\lxnote{looks like each vuln needs 2 commits to fix?}
It is worth noting that while items in our dataset are all security patches, some of them are not conventionally technical vulnerabilities but more like security enhancements, such as upgrading weak crypto algorithms to strong ones. 
In this paper, we do not distinguish them.

In Table~\ref{tab:repo_info}, we also list the metadata of each blockchain project.
We can see that Bitcoin and Ethereum contribute 77.8\% of the vulnerabilities in our dataset, whereas the percentages of Monero and Stellar vulnerabilities are relatively low.
This is mainly because Bitcoin and Ethereum have much more code commits than the other two blockchains, holding a similar percentage (76.9\%) of the entire 85,164 commits.
Additionally, we notice that Stellar 
%has a much higher patch-vulnerability ratio
%\lxnote{not easy to see the ratios; may be clearer to add a column in Table 2 to show the ratios.}
has around the same number of patches as Monero, whereas the number of issues/PRs is three times lower (56 v.s. 178).
The main reason is that Stellar developers tended to use one PR to cover multiple-bug fixes at the early stage of Stellar development.

%Next, based on this unique dataset, we perform a three-level vulnerability analysis in \mysec\ref{sec:categorize}, \mysec\ref{sec:cluster}, and \mysec\ref{sec:detect}, respectively.
Based on this unique dataset, we perform a comprehensive vulnerability analysis at three different levels in \mysec\ref{sec:categorize}, \mysec\ref{sec:cluster}, and \mysec\ref{sec:detect}. %, respectively.

% !TeX root = main.tex

% \vspace{-1.5ex}
\section{File-level Module Categorization}
\label{sec:categorize}
% \vspace{-0.5ex}

%While the four target blockchain systems vary in the system designation and implementation, we aim to provide a generalized top-down architecture and further identify the easily-affected components in this architecture. 
%These pieces of information can be a helpful reference for system developers and security analysts as they can handle the common problem-prone components with caution.

At the first-level of our study, we perform the module analysis of patched files. 
We first propose a lightweight method for categorizing vulnerable modules in \mysec\ref{subsec:cat_method}, and then present the categorization result and its implication in \mysec\ref{subsec:vuln_modules}.

% \vspace{-1.5ex}
\subsection{Identifying and Correlating Module Paths for Vulnerable Module Categorization}
\label{subsec:cat_method}
% \vspace{-0.5ex}

%To perform module categorization, a basic idea is to inspect the (patch) file names and their paths corresponding to those vulnerabilities.
We found that 1,037 vulnerable issues/PRs (or more precisely, 2,317 patch commits) totally generated 2,362 unique file paths (544 in Bitcoin, 1,376 in Ethereum, 251 in Monero, and 191 in Stellar), which makes inspecting each individual file time-consuming.
Therefore, we propose to identify the \textit{module path}, i.e., the folder name that could summarize the module of enclosed files (e.g., the ``rpc/'' folder indicates the RPC module).
For some paths of generic names (e.g., the ``src/'' folder), we consider its sub-folders as module paths.
Since Ethereum's folder structure is more complicated than the other three projects, we also consider three additional folders (the ``core/'', ``swarm/'', and ``eth/'' folders) as generic, and consider their sub-folders as module paths.
Eventually, we obtain a total of 146 module paths (28 in Bitcoin, 71 in Ethereum, 26 in Monero, and 21 in Stellar) from 2,317 patch commits in the four studied blockchains.
%This significantly reduces the workload. %TODO of categorizing vulnerability modules.

Further, since different blockchains have different path names for the same module (e.g., the Consensus module of Bitcoin/Ethereum is in ``consensus/'' while that of Stellar is in ``src/scp/''), we need to correlate those module paths \emph{across projects}.
Our solution is to identify a reference blockchain architecture and map different module paths into this architecture. 
Since many blockchains are based on Bitcoin, we use Bitcoin Core's architecture~\cite{BitcoinCore} as our reference.
For easier understanding, we separate the entire architecture into four layers~\cite{BitcoinCoreOverview}, as shown in \myfig\ref{fig:vuln_modules}, and unify the traditional Miner, Mempool, and Validation Engine components into the Consensus module.
We then manually map those 146 module paths into our blockchain architecture one by one. 
%TODO
% --- a detailed mapping between four blockchains' module paths and their unified modules will be released together with our blockchain vulnerability dataset.
%TODO If a vulnerable issue/PR has no affected module, we categorize it into the \verb|Others| module. 

It is worth noting that a vulnerable issue/PR may affect multiple modules, so the sum of the numbers of vulnerabilities of all the modules is larger than 1,037.
Also, 
%During our categorization, we found that 
some patch commits change only the files directly under the generic ``src/'' folder and do not have module paths.
%As a result, their corresponding vulnerabilities are not mapped into any module.
%To address this, 
We inspect all such patch files (107 in Bitcoin, 31 in Ethereum, 6 in Monero, and 4 in Stellar) and map their corresponding vulnerabilities into the modules in \myfig\ref{fig:vuln_modules} based on the patch file names.\linebreak

% \vspace{-5ex}
%\subsection{Susceptible Blockchain Modules and \red{The Implication Why They are More Risky}}
\subsection{Susceptible Blockchain Modules}
\label{subsec:vuln_modules}
% \vspace{-0.5ex}

\myfig\ref{fig:vuln_modules} shows the result of our module categorization in a layered map of blockchain modules and the numbers of vulnerabilities in those modules.
We can see that modules in the Policy, Peer, Network layers each introduce around one-fourth of the vulnerabilities, while the UI modules and other uncategorized modules contribute the remaining 30\%. 
Among all modules, we find that some modules are more susceptible than the others.
Notably, the modules related to Consensus, Wallet, and NetConn 
%are highly susceptible, \lxnote{may not be good to say so, as it may also depend on ``density'' of vulnerable code.}
contain over 200 issues each.
%\lxnote{vulnerability density (e.g., number of vulnerabilities per thousand lines of code can be more interesting than the absolute numbers of vulnerabilities. If such data is available now, may add into the text here; or, leave it for future.}
Other modules about RPC, GUI/CMD, and Storage are also susceptible, affecting around 100 issues each.
%we find several susceptible modules that have much more vulnerabilities than others, including the Consensus module in the Policy layer, the Wallet and Storage modules in the Peer layer, the NetConn and RPC modules in the Network layer, and the GUI/CMD module in the UI layer.
%We now highlight these modules in a bottom-up order:
We observe that:

\begin{figure}[t!]
  \centering
%  \vspace{-2ex}
  \includegraphics[width=0.8\linewidth]{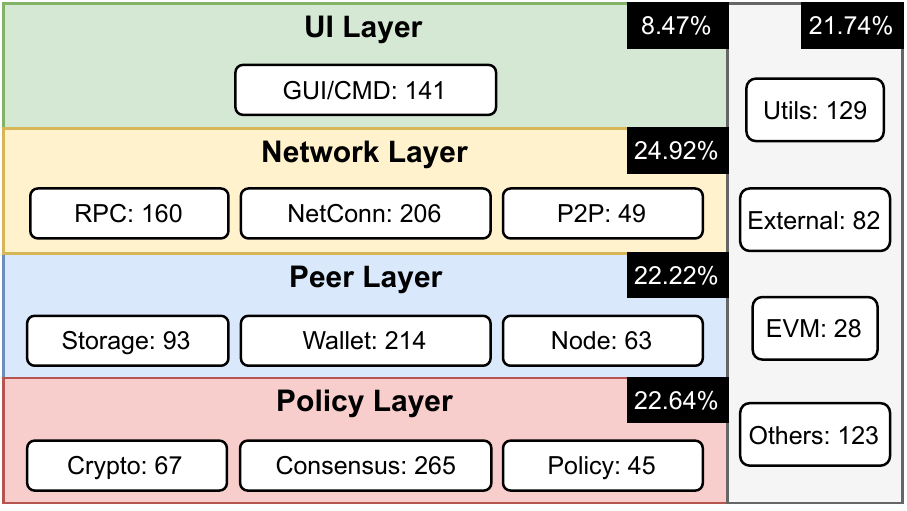}
  % \vspace{-3ex}
  \caption{A layered map of blockchain vulnerabilities in different modules.}
  \label{fig:vuln_modules}
  % \vspace{-5ex}
\end{figure}

\begin{itemize}[leftmargin=1em,nosep,after=\vspace{\baselineskip}]
  \item 
    The Consensus module covers the consensus (e.g., the Proof-of-Work mechanism~\cite{nakamoto2019bitcoin}), miner, block/transaction related components.
    Unfortunately, it was affected by 265 vulnerabilities, with the major module path from the ``consensus/'' folder.
    Other module paths include ``miner/'', ``ethchain/'', ``src/cryptonote\_core/'', ``src/scp/'', and ``src/ledger/''.
    % The vulnerabilities that fall in this module are usually related to transaction verification or block validation, which may lead to serious security issues.

  \item 
    In the Peer layer, the Wallet module handles transactions for each peer and the Storage module manages the storage of those transactions.
    As shown in \myfig\ref{fig:vuln_modules}, the Wallet module was affected by 214 vulnerabilities, which are mainly from the ``src/wallet/'' and ``accounts/'' module paths.
    In contrast, the Storage was affected by 93 vulnerabilities, all of which are from database-related module paths, such as ``src/blockchain\_db/'', ``src/leveldb/'', and ``ethdb/''. 

  \item 
    The NetConn and RPC modules collectively incurred the most blockchain vulnerabilities in our dataset.
    As a distributed system by nature, blockchain systems heavily rely on network synchronization and RPC (Remote Procedure Call).
    Since it deals with complex network communication of different peers, multiple security issues could occur, such as data race, deadlock, resource leak, and denial-of-service. 

  \item 
    Surprisingly, the GUI/CMD module is also a major source, with 141 vulnerabilities from the module paths like ``src/qt/'', ``ethereal/ui/'', ``src/daemon/'', and ``cmd/''.
    The underlying faults vary, but segfault and deadlock are typical bugs.

\end{itemize}

\section{Text-level Type Clustering}
\label{sec:cluster}
% \vspace{-0.5ex}

At the second-level of our study, we conduct the type analysis by analyzing vulnerability text.
%The goal of our text-level vulnerability analysis is to find out the common and distinct vulnerability types of blockchain systems. 
In this section, we first present a NLP-based approach for clustering vulnerability types in \mysec\ref{subsec:cluster_method}, and then summarize the clustering results and showcase common blockchain vulnerability types in \mysec\ref{subsec:vuln_types}, including the ones not known before.

% \vspace{-1.5ex}
\subsection{NLP-based Analysis of Vulnerability Titles for Type Clustering}
\label{subsec:cluster_method}
% \vspace{-0.5ex}

\begin{table*}[t!]
% \vspace{-2ex}
  \caption{Examples of the cleaned issue/PR titles and their corresponding type keywords extracted.}
\label{tab:title_example}
% \vspace{-3ex}
\begin{adjustbox}{center} %angle=90
\scalebox{0.87}{
	\begin{tabular}{|c|l|l|l|}
			\hline
            \textbf{ID} &	\textbf{Raw Title}                    & \textbf{Cleaned Title}                      & \textbf{Type Keywords}      \\ \hline
            \textbf{E1} &	accounts: fix two races in the account manager         & {[}`fix', `two', `races', `in', `the', `account', `manager'{]}                         & {[}`two', `races'{]} \\
            \textbf{E2} &	blockchain\_db: sanity check on tx/hash vector sizes & {[}`sanity', `check', `on', `transaction', `hash', `vector', `sizes'{]}       & {[}`sanity', `check'{]}       \\
            \textbf{E3} &    {[}net{]} Avoid possibility of NULL pointer dereference & {[}`avoid', `null', `pointer', `dereference'{]} & {[}`null', `pointer', `dereference'{]} \\
            \textbf{E4} &	wallet: Fix uninitialized read in bumpfee(…)         & {[}`fix', `uninitialized', `read', `in', `bumpfee'{]}                         & {[}`uninitialized', `read'{]} \\
            \textbf{E5} &	Prevent DOS attacks on in-flight data structures     & {[}`prevent', `dos', `attacks', `on', `in', `flight', `data', `structures'{]} & {[}`dos', `attacks'{]}        \\

			%avoid overflow while computing Price for database    & {[}`avoid', `overflow', `while', `computing', `price', `for', `database'{]}   & {[}`overflow'{]}              \\
            \hline
	\end{tabular}%
}
\end{adjustbox}
% \vspace{-2ex}
\end{table*}

\begin{figure}[t!]
	\centering
	\includegraphics[width=0.65\linewidth]{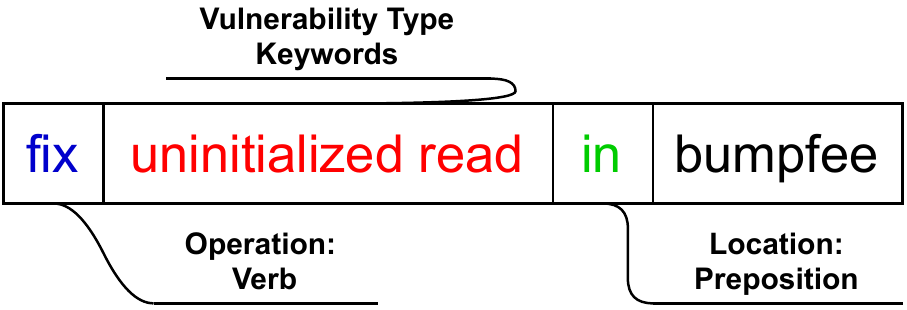}
    % \vspace{-2ex}
    \caption{An example issue/PR title to illustrate the grammatical pattern of vulnerability titles we observed.}
    % \vspace{-2ex}
	\label{fig:keywords}
\end{figure}

%TODO: \blue{(note: issue/PR numbers)}
We find that a vulnerability type is typically captured by the title of an issue/PR page, e.g., Bitcoin PR \#17640 ``wallet: Fix uninitialized read in bumpfee(…),'' where ``uninitialized read'' is the type.
However, simply clustering issue/PR titles does not generate good-quality clusters about vulnerability types because each title could have some noises.
For instance, in the earlier example, ``wallet'' and ``bumpfee'' would affect the clustering quality.
To address this problem, we propose a novel NLP-based method to first extract \textit{type keywords} before we conduct actual clustering. 
This method is based on a grammatical pattern of vulnerability titles we observed, that a type is often a noun phrase
located in between a verb (e.g., ``fix'') and a preposition (e.g., ``in'').
\myfig\ref{fig:keywords} shows an intuitive illustration. 
Overall, our approach consists of two major steps: NLP-based keyword extraction and clustering the obtained type keywords.
%\lxnote{it'd be clearer to state how this clustering is different from the clustering used for keyword-based filtering in Section 3.2.}
Before these two steps, we also need to perform some pre-processing.

\textbf{Pre-processing.}
%Before applying the NLP analysis for keyword extraction, we perform some pre-processing of issue/PR titles so that they are cleaned for next-stage analysis.
To this end, we remove useless words and formalize remaining words in the vulnerability titles.
Specifically, the useless words include
(i) the module/version information (e.g., the word before ``:'', such as the ``wallet'' above, or the word inside ``[]'', such as ``[rpc]'' or ``[RELEASE]''), 
(ii) the special word (e.g., ``SEC-*'' for Ethereum and one-character word like ``a''; note that numbers and symbols like ``--'' or ``(...)'' could be automatically handled by tokenizing),
and (iii) noun-like adjective words (e.g., ``possibility of'' and ``use of'').
After cleaning useless words, we further formalize the remaining words by setting them to the lower case and tokenizing them via the NLP \texttt{nltk}~\cite{nltk} library's \texttt{RegexpTokenizer}.
During this process, we also unify a few words (e.g., replacing all ``tx''/``txs''/``txns'' using ``transaction'').
In Table~\ref{tab:title_example}, we list several example titles our script automatically cleaned.

%In the case of the example shown in \myfig\ref{fig:keywords}, this issue/PR meant to fix the use of an uninitialized parameter \verb|old_fee| in the RPC function \verb|bumpfee()|.
%The vulnerability type \verb|uninitialized read| locates between the Verb \verb|fix| and the Preposition \verb|in|.
%It is safe to assume that the type of vulnerability is usually a Noun or Phrase, such as `deadlock,' `overflow,' `off-by-one,' and `denial-of-service,' which locates between a Verb (i.e., the operation on the vulnerability) and a Preposition (i.e., just before the location or purpose).
%After checking this law of vulnerability issue/PR titles, we find it matches most of the cases in our dataset.

\textbf{NLP-based keyword extraction.}
According to the grammatical pattern shown in \myfig\ref{fig:keywords}, our objective is to find the \textit{target} verb and preposition that could determine the range of type words.
However, one vulnerability title may contain multiple verbs or prepositions.
Moreover, some verbs mainly act as nouns in our context, such as ``check'' and ``leak''.
%\lxedit{leading to poor performance in standard part-of-speech tagging libraries (e.g., \texttt{nltk}~\cite{nltk} library's \texttt{pos\_tag()}).
% Not verb for leak, is, do (not), using, check (not checking), missing, 
Based on these two reasons, 
we do not directly use the \texttt{nltk}~\cite{nltk} library's \texttt{pos\_tag()} for a real-time part-of-speech analysis.
Instead, 
%we build an additional vocabulary of verbs and prepositions and count their frequencies in our dataset, which can be used to facilitate the identification of vulnerabilities types after performing a part-of-speech analysis of the words in our cleaned vulnerability titles.
we perform a \textit{pre}-analysis of words' parts of speech in our cleaned vulnerability titles and build a vocabulary of verbs and prepositions and count their frequencies in our dataset.
%}
Eventually, we obtain a list of 33 verbs and 21 prepositions and rank them by frequencies.
Table~\ref{tab:pos_example} shows the top 10 frequently used verbs and prepositions in our dataset.

Based on our vocabulary of verbs and prepositions and their frequencies, we are able to automatically locate the target verb and preposition for a \textit{cleaned} vulnerability title in various situations using the following rules:

\begin{itemize}[leftmargin=1em,nosep,after=\vspace{\baselineskip}]
    \item If only one verb and one preposition exist and the preposition appears after the verb (with one or more words in between), such a verb and preposition, e.g., the word \texttt{fix} and \texttt{in} of the example E1 in Table~\ref{tab:title_example}, are the target words.

	\item If there is no verb but the preposition exists (e.g., the example E2) or there is no preposition but the verb exists (e.g., the example E3), the preposition or the verb will be determined as the target.
	%, respectively.

    \item If multiple verbs appear in a title, the one with the \textit{highest} frequency will be regarded as the target verb.
      For example, in \myfig\ref{fig:keywords} (or the example E4), the word \texttt{fix} has higher frequency than the word \texttt{read} in our vocabulary, \texttt{fix} is used as the target verb.

    \item If multiple prepositions appear in a title, the \textit{first} one appearing after the target verb (with one or more words in between) is determined as the target preposition. 
      For instance, in the example E5 in Table~\ref{tab:title_example}, both words \texttt{on} and \texttt{in} are prepositions, but since the word \texttt{on} appears before \texttt{in}, \texttt{on} is then determined as the target preposition.

	\item If none of above applies for a vulnerability title, we conclude that it has no target word. 
\end{itemize}

\begin{table}[t!]
  \caption{The top 10 frequently used verbs and prepositions.} % in our dataset of vulnerability titles.}
% \vspace{-2ex}
\label{tab:pos_example}
\begin{adjustbox}{center}
	\scalebox{0.9}{
		\begin{tabular}{|c|ccccc|}
			\hline
			\multirow{2}{*}{\textbf{Verb}}        & add   & remove  & fix    & make  & fixed    \\
			                                      & set   & avoid   & improve& handling & added \\ \hline
			\multirow{2}{*}{\textbf{Preposition}} & in    & for     & on     & of    & with   \\
		                                      & from  & by      & before & if    & after  \\ \hline
		\end{tabular}%
	}
\end{adjustbox}
% \vspace{-4ex}
\end{table}

After recognizing the target verb and preposition for each vulnerability title, the keywords in between the two target words are extracted as the type for the vulnerability.
However, as we list above, some cleaned titles may end up with only one target word or even no any target word.
We handle those special titles as follows:
\begin{itemize}[leftmargin=1em,nosep,after=\vspace{\baselineskip}]
\item If only the target verb exists, all words after the target verb will be regarded as the type keywords.
\item If only the target preposition exists, all words before the target preposition will be treated as the type keywords.
\item If no target word exists, the entire cleaned title becomes the type keywords.
\end{itemize}

\textbf{Clustering type keywords.}
With the extracted type keywords, we aim to cluster them based on their semantic meaning rather than their appearance as a string of letters. 
Thus, after embedding all the keywords into the vector space using word2vec~\cite{mikolov2013wordrepresent}, we choose the Word Mover's Distance (WMD)~\cite{kusner2015wmd} as the similarity metric.
Another reason for applying WMD is that it performs well on short sentences like our type keywords.
Then, we calculate their pairwise similarity with WMD and generate a large similarity matrix.

The last step is to cluster the type keywords based on the similarity matrix.
To reach an optimal clustering result, we tested four clustering algorithms: K-means~\cite{arthur2007kmeans}, Gaussian Mixture~\cite{gaucluster}, Agglomerative Clustering~\cite{aggcluster}, and Affinity Propagation (AP)~\cite{frey2007affinity}.
The first three algorithms require a pre-defined number of clusters as the key parameter, while AP needs a damping factor.
For the first three algorithms, we tried a wide range of cluster numbers from 25 to 225 with an interval of 2.
For AP, we tried the damping factor from 0.5 to 1 with an interval of 0.01.
We kept other parameters unchanged as default.
After clustering with the given parameters, we computed the Silhouette Coefficient score~\cite{rouss1987silhouette} to determine the performance of the corresponding combination.
As a result, Agglomerative clustering with 125 clusters was the best setting for our similarity matrix, which reached a coefficient score of 0.66.

\subsection{Common Blockchain Vulnerability Types}
\label{subsec:vuln_types}
% \vspace{-0.5ex}

According to Table~\ref{tab:vuln_types}, we obtain not only the traditional vulnerabilities, such as race condition and sanity check, but also blockchain-specific vulnerabilities.
Among the top 20 vulnerability types, we find that seven of them are related to blockchains' characteristics.
In particular, the 130 (22.1\%) vulnerabilities from four types (T4, T7, T9, and T12) are blockchain-specific, which are related to blockchains' transaction, block, peer/node, and wallet key/password.
Additionally, we have three more vulnerability types, T2, T14, and T20, that have some portions of their vulnerabilities related to blockchains' features.
The rest of 366 (62.4\%) vulnerabilities are solely the traditional vulnerabilities, not specific to blockchains.

%In the next three paragraphs, we explain three categories of these blockchain types:
Next, we explain three categories of these blockchain types:
\textit{specific}, \textit{partially specific}, and \textit{traditional}.
For the patterns of blockchain-specific vulnerabilities, we will present them in \mysec\ref{subsec:code_pattern}.
%Here we focus on the types only.

\textbf{Blockchain-specific vulnerability types.}
Since transactions, blocks, gas fees are the unique characteristics of blockchain systems, the type T4 and T7 record a large number of such new vulnerabilities.
Examples are Bitcoin PR \#8312 ``Fix mempool DoS vulnerability from malleated transactions'' and Ethereum PR \#1354 ``gpo non-existent block checks''.
Moreover, as a peer-to-peer software by nature, blockchains could suffer from peer/node vulnerabilities.
By inspecting 28 such vulnerabilities in the type T9, we find that they are mainly related to the unique P2P features in blockchains, such as header sync and block validation.
Examples include Bitcoin PR \#10345 ``timeout for headers sync'' and Ethereum issue \#604 ``SEC-41 Peer TD in NewBlockMsg not verified''.
Lastly, blockchain systems often provide wallets to end users, which cause the new vulnerabilities related to wallet keys and passwords in the type T12.
For example, Bitcoin PR \#10308 describes the vulnerability patch of ``[wallet] securely erase potentially sensitive keys/values''.

\begin{table}[t!]
\caption{The top 20 blockchain vulnerability types that affect at least ten vulnerabilities in our dataset.}
% \vspace{-3ex}
\label{tab:vuln_types}
\begin{adjustbox}{center} %angle=90
\scalebox{0.85}{
\begin{threeparttable}
		\begin{tabular}{|c|c|ccccc|c|}
			\hline
			\multirow{2}{*}{\textbf{ID}} &
			  \multirow{2}{*}{\textbf{Type}} &
			  \multicolumn{5}{c|}{\textbf{\# Vulnerability Issues/PRs}} &
			  \multirow{2}{*}{\textbf{Specific?}*} \\ \cline{3-7}
			 &
			   &
			  \multicolumn{1}{c|}{\textbf{All}} &
			  \textbf{B}$^{\diamond}$ &
			  \textbf{E}$^{\diamond}$ &
			  \textbf{M}$^{\diamond}$ &
			  \textbf{S}$^{\diamond}$ &
			   \\ \hline
			\textbf{T1}  & Race Condition      & \multicolumn{1}{c|}{77} & 14 & 48 & 10 & 5 & --    \\
			\textbf{T2}  & Check/Validation    & \multicolumn{1}{c|}{64} & 36 & 14 & 10 & 4 & \ling \\
			\textbf{T3}  & Resource Leak       & \multicolumn{1}{c|}{47} & 24 & 12 & 9  & 2 & --    \\
			\textbf{T4}  & Transaction Related & \multicolumn{1}{c|}{43} & 24 & 9  & 6  & 4 & \gou  \\
			\textbf{T5}  & Deadlock            & \multicolumn{1}{c|}{36} & 16 & 13 & 6  & 1 & --    \\
			\textbf{T6}  & Go Panic            & \multicolumn{1}{c|}{36} & 0  & 36 & 0  & 0 & --    \\
			\textbf{T7}  & Block Related       & \multicolumn{1}{c|}{34} & 9  & 21 & 4  & 0 & \gou  \\
			\textbf{T8}  & Denial-of-Service   & \multicolumn{1}{c|}{31} & 17 & 11 & 3  & 0 & --    \\
			\textbf{T9}  & Peer/Node Related   & \multicolumn{1}{c|}{28} & 12 & 11 & 3  & 2 & \gou  \\
			\textbf{T10} & Sanity Check        & \multicolumn{1}{c|}{28} & 11 & 3  & 13 & 1 & --    \\
			\textbf{T11} & Overflow            & \multicolumn{1}{c|}{27} & 11 & 8  & 6  & 2 & --    \\
			\textbf{T12} & Wallet Key/Password & \multicolumn{1}{c|}{25} & 12 & 6  & 7  & 0 & \gou  \\
			\textbf{T13} & Uninitialized Read  & \multicolumn{1}{c|}{19} & 14 & 0  & 5  & 0 & --    \\
			\textbf{T14} & RPC Related         & \multicolumn{1}{c|}{16} & 9  & 5  & 2  & 0 & \ling \\
			\textbf{T15} & Out-of-Bound        & \multicolumn{1}{c|}{14} & 9  & 4  & 1  & 0 & --    \\
			\textbf{T16} & Off-by-One          & \multicolumn{1}{c|}{14} & 5  & 2  & 7  & 0 & --    \\
			\textbf{T17} & Segfault            & \multicolumn{1}{c|}{13} & 13 & 0  & 0  & 0 & --    \\
			\textbf{T18} & Memory Pool         & \multicolumn{1}{c|}{12} & 10 & 1  & 1  & 0 & --    \\
			\textbf{T19} & Nil Pointer Deref   & \multicolumn{1}{c|}{12} & 6  & 5  & 1  & 0 & --    \\
			\textbf{T20} & Database Corruption & \multicolumn{1}{c|}{11} & 4  & 3  & 4  & 0 & \ling \\ \hline
			\textbf{Sum} &
			  \multirow{2}{*}{--} &
			  \multicolumn{1}{c|}{\textbf{587}} &
			  \textbf{256} &
			  \textbf{212} &
			  \textbf{98} &
			  \textbf{21} &
			  \multirow{2}{*}{--} \\
			\textbf{(\%)} &
			   &
			  \multicolumn{1}{c|}{\textbf{56.6}} &
			  \textbf{24.7} &
			  \textbf{20.4} &
			  \textbf{9.5} &
			  \textbf{2.0} &
			   \\ \hline
		\end{tabular}%
\begin{tablenotes}
\item \small *: \gou~means \textit{most} in this type are blockchain-specific and \ling~means \textit{some} are specific.
  %This column indicates whether vulnerabilities in one type are blockchain-specific, where \gou~means \textit{most} vulnerabilities in this type are blockchain-specific and \ling~means \textit{some} are blockchain-specific.
\item \small $^{\diamond}$: B, E, M, and S represent Bitcoin, Ethereum, Monero, and Stellar, respectively.
\end{tablenotes}
\end{threeparttable}
}
\end{adjustbox}
% \vspace{-4ex}
\end{table}

\textbf{Partially blockchain-specific vulnerability types.}
We also observe three vulnerability types partially specific to blockchains, i.e., T2, T14, and T20.
Specifically, 64 vulnerabilities in the type T2 performed various checks, e.g., error and length checks, and some of them checked blockchain-related properties.
For example, Bitcoin issue \#1167 ``check for duplicate transactions earlier'' for DoS prevention, and Ethereum PR \#20546 ``check propagated block malformation on receiption''.
In contrast, the type T14 and T20 fixed more traditional vulnerabilities related to RPC calls and database corruption (due to exceptional closing), with a few vulnerabilities directly related to blockchains.
Examples of blockchain-related vulnerabilities are Ethereum PR \#19401 ``implement cli-configurable global gas cap for RPC calls'' and Monero issue\#706 ``DB corruption'' due to unfinished blockchain tasks.

\textbf{Traditional vulnerability types in blockchains.}
Besides\linebreak blockchain-specific vulnerabilities, Table~\ref{tab:vuln_types} also shows that 366 vulnerabilities are solely from the 13 traditional vulnerability types.
The top types, such as race condition, deadlock, and denial-of-service, are more frequent probably because it is difficult for blockchain systems to avoid them due to the sync among distributed nodes.

%Besides the overall distribution of the top 20 vulnerability types in our dataset, 
\textbf{Further analysis.}
According to the detailed distribution of vulnerability types across different blockchain projects in Table~\ref{tab:vuln_types}, we make three observations.
First, Ethereum has more than half of the T1 (Race) vulnerabilities, much higher than the other three.
After investigating all the race-related vulnerability issues/PRs, we identify that the Swarm~\cite{swarm} subsystem is the major cause.
Specifically, Swarm is only available in Ethereum and used for distributed storage and content distribution. %TODO (e.g., node-to-node messaging, media streaming, decentralized database services).
Second, we notice that T6 (Go Panic) appears only in Ethereum because only Ethereum is implemented in Go.
Moreover, since Go is a memory-safe language, Ethereum has fewer memory-related (T13, T18) vulnerability issues/PRs than Bitcoin.
Third, we find that Monero has the most number of T10 (Sanity Check) and T16 (Off-by-One) vulnerabilities, while Stellar has the least number of vulnerability types since it is relatively new.

\section{Code-level Pattern Analysis}
\label{sec:detect}
% \vspace{-0.5ex}

\begin{table*}
	\centering
 % \vspace{-1ex}
	\caption{The evolution from raw code hunks to their code fragments and code change signatures.}
 % \vspace{-2ex}
	\label{tab:raw_code_hunk}
	\sbox{\bigimage}{
		\begin{subtable}[b]{0.49\textwidth}
			\centering
			\caption{Example 1: Monero commit 1d5e8f46.}
    		% \vspace{-1ex}
		    \label{tab:exp_1}
		    % \vspace{0pt}
		    % \resizebox{\linewidth}{!}{
		    \begin{adjustbox}{center} %angle=90
			\scalebox{0.76}{
			    \begin{tabular}{rcl} 
					\hline
					\multicolumn{3}{l}{{\cellcolor[rgb]{0.925,0.957,1}}\begin{tabular}[c]{@{}>{\cellcolor[rgb]{0.925,0.957,1}}l@{}}src/crypto/tree-hash.c - void tree\_hash(\\const char (*hashes)[HASH\_SIZE], size\_t count, char *root\_hash) \{ \end{tabular}}  \\ 
					\hline
					\multicolumn{3}{l}{\textbf{Code Hunk} (line 1--23)}                                                                                                                                                                                                                            \\ 
					\hline
					1                                                    &   & \texttt{size\_t cnt = tree\_hash\_cnt( count );}                                                                                                                                                                     \\
					\rowcolor[rgb]{0.996,0.863,0.855} 2                  & - & \texttt{char ints[cnt][HASH\_SIZE];}                                                                                                                                                                                 \\
					\rowcolor[rgb]{0.996,0.863,0.855} 3                  & - & \texttt{memset(ints, 0 , sizeof(ints)); // zero out as extra...}                                                                                                                                                     \\
					\rowcolor[rgb]{0.8,0.996,0.796} 4                    & + & \texttt{char *ints = calloc(cnt, HASH\_SIZE); // zero out as extra...}                                                                                                                                               \\
					\rowcolor[rgb]{0.8,0.996,0.796} 5                    & + & \texttt{assert(ints);}                                                                                                                                                                                               \\
					6                                                    &   & \texttt{memcpy(ints, hashes, (2 * cnt - count) * HASH\_SIZE);}                                                                                                                                                       \\
					7                                                    &   & \texttt{for (\{OMIT\}) \{}                                                                                                                                                                                            \\
					\rowcolor[rgb]{0.996,0.863,0.855} 8                  & - & \texttt{cn\_fast\_hash(hashes[i], 64, ints[j]);}                                                                                                                                                                     \\
					\rowcolor[rgb]{0.8,0.996,0.796} 9                    & + & \texttt{cn\_fast\_hash(hashes[i], 64, ints + j * HASH\_SIZE);}                                                                                                                                                       \\
					10-14                                                &   & \texttt{...}                                                                                                                                                                                                          \\
					\rowcolor[rgb]{0.996,0.863,0.855} 15                 & - & \texttt{cn\_fast\_hash(ints[i], 64, ints[j]);}                                                                                                                                                                       \\
					\rowcolor[rgb]{0.8,0.996,0.796} 16                   & + & \texttt{cn\_fast\_hash(ints + i * HASH\_SIZE, 64, ints + j * HASH\_SIZE);}                                                                                                                                           \\
					17-18                                                &   & \texttt{\}\}}                                                                                                                                                                                                          \\
					\rowcolor[rgb]{0.996,0.863,0.855} 19                 & - & \texttt{cn\_fast\_hash(ints[0], 64, root\_hash);}                                                                                                                                                                    \\
					\rowcolor[rgb]{0.8,0.996,0.796} 20                   & + & \texttt{cn\_fast\_hash(ints, 64, root\_hash);}                                                                                                                                                                       \\
					\rowcolor[rgb]{0.8,0.996,0.796} 21                   & + & \texttt{free(ints);}                                                                                                                                                                                                 \\
					22                                                   &   & \texttt{\}}                                                                                                                                                                                                          \\
					\hline
					\multicolumn{3}{l}{\textbf{Code Fragments}}                                                                                                                                                                                                             \\ 
					\hline
					\multicolumn{1}{c|}{\multirow{4}{*}{\textbf{F1-1} }} & - & \texttt{char ints[cnt][HASH\_SIZE];}                                                                                                                                                                                 \\
					\multicolumn{1}{c|}{}                                & - & \texttt{memset(ints, 0 , sizeof(ints));}                                                                                                                                                                             \\
					\multicolumn{1}{c|}{}                                & + & \texttt{char *ints = calloc(cnt, HASH\_SIZE);}                                                                                                                                                                       \\
					\multicolumn{1}{c|}{}                                & + & \texttt{assert(ints);}                                                                                                                                                                                               \\ 
					\hline
					\multicolumn{1}{c|}{\multirow{2}{*}{\textbf{F1-2} }} & - & \texttt{cn\_fast\_hash(hashes[i], 64, ints[j]);}                                                                                                                                                                     \\
					\multicolumn{1}{c|}{}                                & + & \texttt{cn\_fast\_hash(hashes[i], 64, ints + j * HASH\_SIZE);}                                                                                                                                                       \\ 
					\hline
					\multicolumn{1}{l|}{\multirow{2}{*}{\textbf{F1-3}}}  & - & \texttt{cn\_fast\_hash(ints[i], 64, ints[j]);}                                                                                                                                                                       \\
					\multicolumn{1}{l|}{}                                & + & \texttt{cn\_fast\_hash(ints + i * HASH\_SIZE, 64, ints + j * HASH\_SIZE);}                                                                                                                                           \\ 
					\hline
					\multicolumn{1}{l|}{\multirow{3}{*}{\textbf{F1-4}}}  & - & \texttt{cn\_fast\_hash(ints[0], 64, root\_hash);}                                                                                                                                                                    \\
					\multicolumn{1}{l|}{}                                & + & \texttt{cn\_fast\_hash(ints, 64, root\_hash);}                                                                                                                                                                       \\
					\multicolumn{1}{l|}{}                                & + & \texttt{free(ints);}                                                                                                                                                                                                 \\ 
					\hline
					\multicolumn{3}{l}{\textbf{Code Change Signatures}}                                                                                                                                                                                                                    \\ 
					\hline
                    \multicolumn{1}{c|}{\textbf{S1-1}}                  & \multicolumn{2}{l}{\texttt{VAR[][] ==\textgreater{} calloc() memset() assert()}}                                                                                                                                         \\ 
					\hline
					\multicolumn{1}{c|}{\textbf{S1-2}}                  & \multicolumn{2}{l}{\texttt{cn\_fast\_hash()}}                                                                                                                                                                            \\ 
					\hline
					\multicolumn{1}{c|}{\textbf{S1-3}}                   & \multicolumn{2}{l}{\texttt{cn\_fast\_hash()}}                                                                                                                                                                            \\ 
					\hline
					\multicolumn{1}{c|}{\textbf{S1-4}}                   & \multicolumn{2}{l}{\texttt{cn\_fast\_hash() free()}}                                                                                                                                                                     \\
					\hline
				\end{tabular}
			}
			\end{adjustbox}	
	    \end{subtable}
	}
	\usebox{\bigimage}\hfill
	\begin{minipage}[b][\ht\bigimage][s]{0.49\textwidth}
		\begin{subtable}{\textwidth}
			\centering
			\caption{Example 2: Ethereum commit b765e2d1.}
    		% \vspace{-1ex}
	    	\label{tab:exp_2}
	    	% \resizebox{\linewidth}{!}{
	    	\begin{adjustbox}{center} %angle=90
			\scalebox{0.76}{
		    	\begin{tabular}{rcl}
					\hline
					\multicolumn{3}{l}{{\cellcolor[rgb]{0.925,0.957,1}}\begin{tabular}[c]{@{}>{\cellcolor[rgb]{0.925,0.957,1}}l@{}} core/transaction\_pool.go - func (pool *TxPool) \\ValidateTransaction(tx *types.Transaction) error \{ \end{tabular}}       \\ 
					\hline
					\multicolumn{3}{l}{\textbf{Code Hunk} (line 1--9)}                                                                                                                                                                                                                 \\ 
					\hline
					1-2                                                &   & \texttt{return fmt.Errorf("tx.v != (28 \textbar{}\textbar{} 27) = \% v", v)\}}                                                                                                                                    \\
					\rowcolor[rgb]{0.8,0.996,0.796} 3                  & + & \texttt{senderAddr := tx.From()}                                                                                                                                                                            \\
					\rowcolor[rgb]{0.8,0.996,0.796} 4                  & + & \texttt{if senderAddr == nil \textbar{}\textbar{} len(senderAddr) != 20 \{}                                                                                                                                 \\
					\rowcolor[rgb]{0.8,0.996,0.796} 5                  & + & \texttt{return fmt.Errorf("invalid sender")}                                                                                                                                                                 \\
					\rowcolor[rgb]{0.8,0.996,0.796} 6                  & + & \texttt{\}}                                                                                                                                                                                                 \\
					7                                                  &   & \texttt{/* XXX this kind of validation needs to happen elsewhere...}                                                                                                                                     \\
					\hline
					\multicolumn{3}{l}{\textbf{Code Fragment}}                                                                                                                                                                                                  \\ 
					\hline
					\multicolumn{1}{c|}{\multirow{3}{*}{\textbf{F2}}} & + & \texttt{senderAddr := tx.From()}                                                                                                                                                                            \\
					\multicolumn{1}{l|}{}                              & + & \texttt{if senderAddr == nil \textbar{}\textbar{} len(senderAddr) != 20}                                                                                                                                    \\
					\multicolumn{1}{l|}{}                              & + & \texttt{return fmt.Errorf("invalid sender")}                                                                                                                                                                 \\ 
					\hline
					\multicolumn{3}{l}{\textbf{Code Change Signature}}                                                                                                                                                                                              \\ 
					\hline
					\multicolumn{1}{c|}{\textbf{S2}}                  & \multicolumn{2}{l}{\texttt{From() if NIL \textbar{}\textbar{} LEN return ERR}}                                                                                                                                  \\ 
					\hline
				\end{tabular}
			}
			\end{adjustbox}
    	\end{subtable}
    	\vfill
    	\vspace*{0.27cm}
    	\begin{subtable}{\textwidth}
    		\centering
			\caption{Example 3: Ethereum commit 7c24cd79.}
		    \label{tab:exp_3}
		    % \resizebox{\linewidth}{!}{
		    \begin{adjustbox}{center} %angle=90
			\scalebox{0.76}{
			    \begin{tabular}{rcl}
					\hline
					\multicolumn{3}{l}{{\cellcolor[rgb]{0.925,0.957,1}}\begin{tabular}[c]{@{}>{\cellcolor[rgb]{0.925,0.957,1}}l@{}}chain/transaction\_pool.go - func (pool *TxPool) \\ValidateTransaction(tx *types.Transaction) error \{ \end{tabular}}    \\ 
					\hline
					\multicolumn{3}{l}{\textbf{Code Hunk} (line 1--10)}                                                                                                                                                                                                                  \\ 
					\hline
					1                                                  &   & \texttt{//sender := pool.\{OMIT\}.proState.GetAccount(tx.Sender())}                                                                                                                                         \\
					\rowcolor[rgb]{0.996,0.863,0.855} 2                & - & \texttt{sender := pool.\{OMIT\}.CurrentState().GetAccount(tx.Sender())}                                                                                                                                     \\
					\rowcolor[rgb]{0.8,0.996,0.796} 3                  & + & \texttt{senderAddr := tx.Sender()}                                                                                                                                                                          \\
					\rowcolor[rgb]{0.8,0.996,0.796} 4                  & + & \texttt{if senderAddr == nil \{}                                                                                                                                                                            \\
					\rowcolor[rgb]{0.8,0.996,0.796} 5                  & + & \texttt{return fmt.Errorf("Invalid sender")}                                                                                                                                                                \\
					\rowcolor[rgb]{0.8,0.996,0.796} 6                  & + & \texttt{\}}                                                                                                                                                                                                 \\
					\rowcolor[rgb]{0.8,0.996,0.796} 7                  & + & \texttt{sender := pool.\{OMIT\}.CurrentState().GetAccount(senderAddr)}                                                                                \\
					8                                                  &   & \texttt{totAmount := new(big.Int).Set(tx.Value)}                                                                                                                                                            \\
					\hline
					\multicolumn{3}{l}{\textbf{Code Fragment}}                                                                                                                                                                                                  \\ 
					\hline
					\multicolumn{1}{c|}{\multirow{5}{*}{\textbf{F3}}}  & - & \texttt{sender := pool.\{OMIT\}.CurrentState().GetAccount(tx.Sender())}                                                                                                                                     \\
					\multicolumn{1}{l|}{}                              & + & \texttt{senderAddr := tx.Sender()}                                                                                                                                                                          \\
					\multicolumn{1}{l|}{}                              & + & \texttt{if senderAddr == nil}                                                                                                                                                                               \\
					\multicolumn{1}{l|}{}                              & + & \texttt{return fmt.Errorf("Invalid sender")}                                                                                                                                                                \\
					\multicolumn{1}{l|}{}                              & + & \texttt{sender := pool.\{OMIT\}.CurrentState().GetAccount(senderAddr)}                                                                                                                                                                \\ 
					\hline
					\multicolumn{3}{l}{\textbf{Code Change Signature}}                                                                                                                                                                                             \\ 
					\hline
                    \multicolumn{1}{c|}{\textbf{S3}}                   & \multicolumn{2}{l}{\texttt{GetAccount() Sender() if NIL return ERR}}                                                                                                                                            \\
					\hline
				\end{tabular}
			}
			\end{adjustbox}
    	\end{subtable}
	\end{minipage}
% \vspace{-2ex}
\end{table*}

%Next, we introduce our approach to generating code change signatures and clustering them.
%Before these two major steps,
%we first clean up code hunks and turn them into fragments, and then align up the changed lines of code in each fragment.
%
%
%\textbf{Cleaning and splitting each code hunk into fragments.}
%%As mentioned in \mysec\ref{subsec:dataset}, we have 2,317 commits with code hunks in our vulnerability dataset.

%While the text-level clustering provides new findings on blockchains' vulnerability types, it does not reveal their code-level patterns.
At the third-level of our study, we perform the pattern analysis by analyzing vulnerability patch code.
In particular, we focus on blockchain-specific vulnerability types (i.e., the seven types mentioned in \mysec\ref{subsec:vuln_types}) since the code patterns of traditional vulnerability types like race condition, deadlock, overflow, and uninitialized read are well-known (e.g.,~\cite{xuthesis, cai2014magiclock, wang2018detect, lu2017unleashing}). % ompracer2020swain, boldyreva2016heap, milburn2017safelnit
In this section, we first propose our approach to summarizing patch code patterns in~\mysec\ref{sec:extractPattern}, and then present blockchain-specific code patterns in~\mysec\ref{subsec:code_pattern}.
%, and further apply the generated patterns for vulnerability detection in~\mysec~\ref{subsec:apply_pattern}.

% !TeX root = main.tex

% \vspace{-2ex}
\subsection{Generating and Clustering Code Change Signatures for Vulnerability Patterns}
\label{sec:extractPattern}
% \vspace{-1ex}

To obtain vulnerability code-level patterns, our objective is to put similar patch code \textit{changes} into the same cluster so that analysts can summarize patterns from each cluster.
To this end, we need an effective representation of code changes so that it keeps important semantic information yet ignores unimportant or noisy information.
We call this representation the \textit{code change signature}.
Table~\ref{tab:raw_code_hunk} illustrates the evolution process from raw code hunks to their code fragments (i.e., contiguous lines of code) and the corresponding code change signatures using three examples.
Taking the code in Table~\ref{tab:exp_2} and ~\ref{tab:exp_3} as an example, both patches check whether the sender of a transaction is valid.
However, if the variable name \texttt{senderAddr} is different, the similarity between their raw code fragment change (i.e., the syntactic changes indicated by F2 and F3) would be low.
To capture the \textit{essential} changes in patch code, we do not use the syntactic changes but their code change signatures like S2 and S3, the details of which will be illustrated during their generation.

%\TODO{\red{======================START}}

Next, we introduce our approach to generating code change signatures and clustering them.
Before these two major steps,
%\lxnote{why not introduce the 4 steps in order?}
we first clean up code hunks and turn them into fragments, and then align up the changed lines of code in each fragment.
%\lxnote{what does 'calculating the change' mean? Do you mean 'measuring' or 'identifying the change', or 'constructing the change signatures', or others?}

%\smallskip
\textbf{Cleaning and splitting each code hunk into fragments.}
%As mentioned in \mysec\ref{subsec:dataset}, we have 2,317 commits with code hunks in our vulnerability dataset.
The raw code hunks we retrieved contain not only \textit{meaningful} diff code but also test code, neighboring context (e.g., in-line and block comments, unchanged code lines, \texttt{\#include} and \texttt{import} statements), and modification of none-code files (e.g., mark-down, JSON, and text files).
Therefore, we first initiate a cleaning process~\cite{AndroVulns19} to keep only the actual diff code hunks and
separate them into individual fragments by continuous `+' and `-' lines. %(in a way similar to~\cite{AndroVulns19})
Taking the code hunk in Table~\ref{tab:exp_1} as an example, it is separated into four code fragments, F1-1 (line 2-5), F1-2 (line 8-9), F1-3 (line 15-16), and F1-4 (line 19-21) after removing the neighboring context lines (i.e., line 1, 6-7, 10-14, 17-18, 22-23, and the comments in line 3 and 4).
%As a result, we obtain 5,772 code fragments separated from the 2,317 code hunks.

%\smallskip
\textbf{Aligning up changed lines of code in each fragment.}
%As you may have noticed, two statements (i.e., line 3 and line 8 in Table~\ref{tab:exp_3}) remain unprocessed in the previous example.
%These two statements are special since line 8 is changed from line 3, not simply added.
%The only difference between these two statements is the parameter of the function \verb|GetAccount()|.
%Therefore, we can abstract the signature of these two statements as this function name.
%Nevertheless, we need more consideration when dealing with this kind of statements (i.e., `-' line change into `+' line) because the two statements may not modify the same function, or even their types are different (e.g., line 2 and line 4 in Table~\ref{tab:exp_1}).
Before we generate each code fragment's change signature based on deleted and added lines in it,
we need
to first pair up the changed lines of code
%, such as line 19 matched with line 20 instead of line 21 in Table~\ref{tab:exp_1}.
since only some code fragments have one-to-one line change (i.e., at most one `-' line and one `+' line).
For example, in Table~\ref{tab:raw_code_hunk},
only the fragments F1-2 and F1-3 have one-to-one line change. %, and the only `-' line can be simply paired up with the only `+' line.
For a multiple-line change in other fragments, we measure the edit distance similarity between each `-' line and all `+' lines and pair the one with the highest similarity.
For instance, line~3 in Table~\ref{tab:exp_3} is paired with line 8 since it has the highest similarity with line 8 as compared with all the other lines. %, and the similarity is greater than 0.5.
However, some lines could be simply deleted or added, causing their similarity with all other lines to be low.
We handle this by \textit{not} pairing the lines with the highest similarity of less than~0.5.
As a result, line 3 in Table~\ref{tab:exp_1} will not be paired with line 5 due to the low similarity.
%\lxnote{some lines are not paired based on the descriptions here, so it may be better to use a more generic word 'aligning up' changed lines of code for the title; while in the paragraph, can continue to use 'pair' for paired '+' and '-' lines.}

%\smallskip
\textbf{Generating the signatures of code changes.}
%Guided by the above examples, we summarize the following procedures to abstract the signature for every statement in a code fragment.
%For simplicity, we use the line number in the raw code hunk as the corresponding statement in the code fragment in the following paragraphs.
After determining the paired lines of code, we extract their syntactic changes~\cite{AndroVulns19}
to generate the signatures with the following alterations:

%\TODO{\red{======================END}}

%\vspace{-2ex}
\begin{itemize}[leftmargin=1em,nosep,after=\vspace{\baselineskip}]
  \item \textit{(Recognizing and marking the type of statements.)}
    We first determine the control-flow statements by six reserved keywords, \texttt{if}, \texttt{for}, \texttt{while}, \texttt{return}, \texttt{throw}, and \texttt{defer}.
    If a control-flow statement is identified, we keep not only their type keyword but also their logical operators, e.g., ``\texttt{||}'' in line 4 in Table~\ref{tab:exp_2}.
	If a statement does not contain any control-flow keyword, we regard it as a function call if it includes a function or an assignment statement if it does not. 
	For example, neither line 2 and line 4 in Table~\ref{tab:exp_1} have a control-flow keyword, but line 4 contains a function \verb|calloc()|, so we regard line 4 as a function call and line 2 as an assignment. %TODO statement.

  \item \textit{(Preserving the name only for a function call.)}
    For function calls, we found that the function name itself is often enough to capture the statement nature despite parameter changes.
    Therefore, in code change signatures, we eliminate the function's parameters and caller variables.
    For example, we eliminate the three parameters of \texttt{cn\_fast\_hash()} in Table~\ref{tab:exp_1} and keep its function name only.
    As a result, it is easy for the generated three signatures (S1-2/3/4) to be in the same cluster.
    The symbols for calling a function vary, including \texttt{.}, \texttt{->}, and \texttt{::}.
    Additionally, if a function is called by another in a statement, we consider the last one as the actual function call of this statement, e.g., \texttt{GetAccount()} in Table~\ref{tab:exp_3}.

  \item \textit{(Abstracting variable names and variable values.)}
    We also abstract variable names and variable values for a more concise signature.
    Specifically, we substitute variable names with the keyword \texttt{VAR}.
    If a variable is an array, we further add one or more \verb|[][]|, such as \verb|VAR[][]| for line 2 in Table~\ref{tab:exp_1}.
    For variable values, we define six keywords for the substitution of different types of values: 
	\verb|NIL| for \verb|nil|, \verb|null|, and \verb|none|; \verb|BOL| for \verb|true| and \verb|false|; \verb|NUM| for numbers; \verb|TXT| for strings;
    \verb|LEN|/\texttt{SIZE} for size-related functions (e.g., \verb|len()|, \verb|length()|, \verb|size()|, and \verb|sizeof()|);
    and \verb|ERR| for error functions.
\end{itemize}
% \vspace{-2ex}

%After all these alterations, we further insert a special change symbol ``\verb|==>|'' if the raw syntactic changes of
%a pair of `-' and `+' line have been abstracted into the signature, e.g., line 2 and 4 in Table~\ref{tab:exp_1}.
%Otherwise, we do not insert the change symbol, as shown by line 8 and 9 in Table~\ref{tab:exp_1} and line 3 and 8 in Table~\ref{tab:exp_3}.
%After processing all the statements in a code fragment, we generate the fragment signature by concatenating each line-based signature.

% !TeX root = main.tex

\begin{table*}[t!]
	\centering
%\vspace{-2ex}
	\caption{21 blockchain-specific patch code patterns obtained from the clustering result of 3,251 code fragments.}
	\label{tab:pattern_result}
% \vspace{-3ex}
	\begin{adjustbox}{center} %angle=90
	\scalebox{0.746}{
		\begin{threeparttable}
			\begin{tabular}{|c|c|l|l|c|}
				\hline
				\textbf{Type} &
				  \textbf{ID} &
				  \textbf{Description} &
				  \textbf{Pattern (in the revised code change signature with some generalizations)} & %with regular expression; using the symbol ``|'' as OR
				  \textbf{Example*} \\ \hline

				\multirow{6}{*}{\begin{tabular}[c]{@{}c@{}}Transaction\\ Related\end{tabular}}
	             %TODO: tx relay bug: Bitcoin 4450, 16557, 15668, 14220, 15759
				 & \textbf{P1} &
				   Check the transaction sender address &
				  \texttt{From|Sender|address() if ==NIL||LEN()!=NUM | IsValid() return ERR()} &
%				  E: \#272 \\ \cline{2-5}       % E: 195, B: 1002, 448 (a bit)
				  E: \#272~\cite{ETH272} \\ \cline{2-5}       % E: 195, B: 1002, 448 (a bit)
				 & \textbf{P2} &
				  Check the size of transactions in a pool &
				  \texttt{GetSerializeSize()|SIZE() if $>$ MAX\_STANDARD\_TX\_SIZE return BOL} &
				  B: \#2273~\cite{BTC2273} \\ \cline{2-5}      % E: 20352; E: 17178 (this is different, for length)
				 & \textbf{P3} &
				  Shuffle the transaction order; otherwise, fingerprinting &
				  \texttt{clear() selected\_coins() shuffle() push\_back()} &
				  B: \#12699~\cite{BTC12699} \\\cline{2-5}      % B: 14897; B: 8408 (a bit)
				 & \textbf{P4} &
				   Prevent the duplicated transaction &
				   \texttt{BOOST\_FOREACH() insert() if SIZE() != SIZE() return DoS()} &
				  B: \#1167~\cite{BTC1167} \\ \cline{2-5}      % B: 6588; TODO B: 915 unspent also BOOST_FOREACH()
				 & \textbf{P5} &
				  Prevent the malformed transaction  &
				  \texttt{if !IsStandardTx() return DoS()} &
				  B: \#8312~\cite{BTC8312} \\ \cline{2-5}      % B: 13452 (a bit similar), E: 420/711
				 & \textbf{P6} &
				  Prevent the double-spent transaction (relay) &
				  \texttt{if RelayableRespend() VAR = BOL} &
				  B: \#4514~\cite{BTC4514} \\                   % B: 4450 (a bit?); B: 915 unspent removed
	             \hline

				\multirow{6}{*}{\begin{tabular}[c]{@{}c@{}}Block \\ Related\end{tabular}}
				 & \textbf{P7} &
				  Validate the new header not from an invalid block &
				  \texttt{if IsValid() while!= insert() return DoS()} &
				  B: \#11531~\cite{BTC11531} \\ \cline{2-5}     % B: 11487, 5078 (a bit)
				 & \textbf{P8} &
				  Check the gas limit in a block header &
				  \texttt{CalcGasLimit() if Cmp() != NUM return ERR()} &
				  E: \#389~\cite{ETH389} \\ \cline{2-5}       % E: 77, 599 (a bit)
				 & \textbf{P9} &
				  Check the block timestamp &
				  \texttt{time() int64() if <} &
				  M: \#5902~\cite{MONERO5902} \\ \cline{2-5}      % E: 1355 (uint64), 1352 (just 1355)
				 & \textbf{P10} &
				  Validate some block fields (number and hash) not null &
				  \texttt{GetBlockByNumber()|Hash() if != NIL} &
				  E: \#1354~\cite{ETH1354} \\ \cline{2-5}     % E: 19744 and 1939 (hash); E: 1354 and 19573 (GetBlockByNumber: 598..); B: 5959 (index); M: 116 (size); E: 600 (usedGas, 1264 a bit)
				 & \textbf{P11} &
				  Do not connect a corrupted block &
				  \texttt{if CorruptionPossible() return AbortNode()|BOL} &
				  B: \#12561~\cite{BTC12561} \\ \cline{2-5}     % B: 3884
				 & \textbf{P12} &
				  Prevent a malformed block to be propagated or forked &
				  \texttt{if CalcUncleHash()!=UncleHash() | if DeriveSha()!=TxHash() break} &
				  E: \#20546~\cite{ETH20546} \\                 % E: 865 (block number)
	             \hline

				\multirow{3}{*}{\begin{tabular}[c]{@{}c@{}}Peer/Node\\ Related\end{tabular}}
				 & \textbf{P13} &
				  Disconnect after the timeout of header synchronization &
				  \texttt{if GetBlockTime() <= GetAdjustedTime()-NUM  return BOL} &
				  B: \#10345~\cite{BTC10345} \\ \cline{2-5}     % B: 5463
				 & \textbf{P14} &
				  Disconnect outbound peers on the invalid chain &
				  \texttt{if GetHash()!=  fDisconnect = BOL} &
				  B: \#11568~\cite{BTC11568} \\ \cline{2-5}     % B: 11446
				 & \textbf{P15} &
				  Drop the remote peer on an invalid or unverified TD &
				  \texttt{if Cmp(BlockTd) != 0} &
				  E: \#604~\cite{ETH604} \\                   % E: 1451
	             \hline

				\multirow{3}{*}{\begin{tabular}[c]{@{}c@{}}Wallet\\ Key/Password\end{tabular}}
				 & \textbf{P16} &
				  Immediately wipe the memory for critical secret keys &
				  \texttt{rct2sk()|MLSAG\_Gen() memwipe()|memory\_cleanse() return} &
				  M: \#4268~\cite{MONERO4268} \\ \cline{2-5}      % B: 10308
				 %& \textbf{P18} &
				 %  increase incorrect password delay &
				 %  \texttt{if size() \textless NUM Sleep()} &
				 %  B: \#798 and \#799 \\ \cline{2-5}
				 & \textbf{P17} &
				  Try to keep the wallet address in testnet or memory &
				  \texttt{generate() if != MAINNET || create\_address\_file ERR()} &
				  M: \#3315~\cite{MONERO3315} \\ \cline{2-5}      % B: 787, E: 19191
				 & \textbf{P18} &
				  Do not skip asking for password when watch-only &
				  \texttt{if ask\_password()} &
				  M: \#4791~\cite{MONERO4791} \\                  % M: 4324 (a bit); E: 17651 & 15950 ?
	             \hline

				 Other Check
				 %& \textbf{P14} &
				 %  Crash due to invalid account address \red{sim to P3, not needed} &
				 %  \texttt{get\_str() --\textgreater address() if IsValid() throw ERR()} &
				 %  B: \#448 \\ \cline{2-5} 
	             & \textbf{P19} &
				  Check the validity of Quorum set &
				  \texttt{if !isQuorumSetSane() ERR() if throw invalid\_argument()} &
				  S: \#2233~\cite{Stellar2233} \\ \hline           % S: 2209, 2125 (a bit)

	            RPC Related &
				  \textbf{P20} &
				  Enforce a gas cap of caller to protect against DoS &
				  \texttt{if := RPCGasCap() if != NIL if Cmp() > NUM Warn()} &
				  E: \#19401~\cite{ETH19401} \\ \hline

				DB Corruption &
				  \textbf{P21} &
				  Avoid corruption due to unfinished blockchain tasks &
				  \texttt{CRITICAL\_REGION\_LOCAL()} &
				  M: \#706~\cite{MONERO706} \\ \hline            % E: 846
			\end{tabular}%

			\begin{tablenotes}
				\item \small *: This column lists one example issue/PR for each code pattern, where B, E, M, and S represent Bitcoin, Ethereum, Monero, and Stellar, respectively.
			\end{tablenotes}

		\end{threeparttable}
	}
	\end{adjustbox}
% \vspace{-2ex}
\end{table*}

\textbf{Clustering code change signatures.}
As mentioned earlier, we cluster code change signatures from the vulnerabilities of blockchain-specific types. %TODO in the beginning of \mysec\ref{sec:detect}
Since the RPC-related and database corruption types (i.e., the type T14 and T20 in Table~\ref{tab:vuln_types}) have only one or two blockchain-specific vulnerabilities, there is no need to cluster their signatures. 
Eventually, our target is 3,251 code fragments from 194 vulnerabilities of the type T2, T4, T7, T9, and T12 (see \mysec\ref{subsec:vuln_types}).
%After generating code change signatures for each code fragment, we further cluster them.
The clustering process is similar to that in \mysec\ref{subsec:cluster_method}.
One difference is that the WMD similarity is no longer applicable because code fragment signatures cannot be mapped to the token-based vector space.
Therefore, we choose the Normalized Levenshtein distance~\cite{TPAMI07Levenshtein} as the metric for calculating the similarity between code fragment signatures.

To find a suitable clustering algorithm here, we also tested the four algorithms in \mysec\ref{subsec:cluster_method}, i.e., K-means, Gaussian Mixture, Agglomerative Clustering, and Affinity Propagation (AP).
For the first three algorithms that require a pre-estimation of the number of clusters, we compute the Silhouette Coefficient score in a wide range of cluster numbers, but the result is not satisfactory.
Therefore, we choose AP as our code clustering algorithm since it does not require pre-setting the number of clusters and performs well with a gradual tuning of the damping factor to 0.78.
Under this setting,
%By clustering the 3,251 code fragments' change signatures via AP,
we eventually obtain a total of 174 clusters for further pattern analysis.

%We further study the complexity of blockchain system vulnerability fixes, i.e., how many lines of code fragments are needed to fix a vulnerability issue/PR?
%For each vulnerability issue/PR, we count the number of lines of all corresponding code fragments, and draw a CDF plot showed in \myfig\ref{fig:code_lines}.
%As we can see, around half of the vulnerability issues/PRs can be fixed within ten lines of code fragments.
%\begin{figure}[]
%	\centering
%	\includegraphics[width=0.85\linewidth]{lines_cdf.pdf}
%    \caption{CDF of \# lines of code fragments per vulnerability.\red{update!}}
%	\label{fig:code_lines}
%\end{figure}

% \vspace{-1.5ex}
%\subsection{Blockchain-specific Patch Code Patterns and Their Usages}
\subsection{Blockchain-specific Patch Code Patterns}
\label{subsec:code_pattern}
% \vspace{-0.5ex}

% Fang: annotate the table
%\input{codeExp}

After clustering code change signatures,
we inspect all the clusters
and generalize the code patterns from them.
Table~\ref{tab:pattern_result} lists the 21 evidently blockchain-specific vulnerability patterns.
They are organized in seven categories by their types (see~\mysec\ref{subsec:vuln_types}), and most check-related patterns have been categorized into the detailed types.

\textbf{Transaction-related patterns.}
We have identified six patterns (P1--P6) related to blockchain transactions.
%\red{As shown in Table~\ref{tab:pattern_result},} 
They check the sender (P1), size (P2), and order (P3) of a transaction, and prevent duplicated (P4), malformed (P5), double-spent (P6) transactions.
Specifically,
\textbf{P1} checks a sender address function to guarantee non-null values with valid lengths.
\textbf{P2} checks the maximum size of transactions allowed in a pool.
Besides the sender and length, the order of transactions could incur privacy risks like fingerprinting if not randomized.
To address this and as the case of Bitcoin \#12699, \textbf{P3} is to \texttt{clear} the original order, \texttt{shuffle} it, and \texttt{push\_back} the new order.
Both \textbf{P4} and \textbf{P5} check the blockchain structure to prevent duplicated or malformed transactions; otherwise, DoS could happen.
\textbf{P6} prevents double-spent transaction relays via \path{Relayable-Respend()}, which was checked in Bitcoin \#4515 and \#4450.

\textbf{Block-related patterns.}
We identify another six patterns (P7--P12) related to blockchain blocks.
As a basic blockchain unit, a block stores multiple transactions and will be appended to the chain according to the consensus protocols.
%However, appending an unverified block may end up polluting the whole chain.
However, vulnerabilities could happen if the header (P7), gas limit (P8), and timestamp (P9) of a new block is invalid, or if some block fields are not null (P10), or if a corrupted (P11) or malformed (P12) block is identified.
Specifically,
\textbf{P7} validates that a newly appended block header is not from an invalid block.
\textbf{P8} checks the gas limit in a block header, where gas is the fee for running smart contracts in Ethereum~\cite{buterin2014next}.
\textbf{P9} checks whether the timestamp in a block is less than the current \texttt{time()}, such as Monero \#5902 and Ethereum \#1355.
\textbf{P10} validates the block fields like number and hash, and guarantees they are not null.
Both \textbf{P11} and \textbf{P12} check the structure of a block to prevent a corrupted or malformed block being connected or forked.

\textbf{Peer/node-related patterns.}
We also identify three patterns (P13--P15) related to peer/node synchronization and validation.
Specifically, \textbf{P13} checks the time of block header synchronization, and if it is timed out, the node would disconnect.
%Table~\ref{tab:p13exp} shows such an example in Bitcoin \#10345, where the timeout is checked via \texttt{GetBlockTime()}. % and \texttt{GetAdjustedTime()}.
Bitcoin \#10345 shows such an example in Bitcoin \#10345, where the timeout is checked via \texttt{GetBlockTime()}. % and \texttt{GetAdjustedTime()}.
A similar case is Bitcoin \#5463 for the block download timeout.
Additionally, \textbf{P14} and \textbf{P15} perform the validation of remote peers and drop them if they fail.
For example, Bitcoin \#11568 and \#11446 in P14 validate the hash of outbound peers via \texttt{GetHash()}.
P15, on the other hand, checks the Ethereum-specific TD (Total Difficulty) field of a peer and guarantees the advertised TD actually deliverable, as in Ethereum \#604 and \#1451.

\textbf{Wallet-related patterns.}
%Wallet serves as a critical component in the blockchain system, as it deals with sensitive information such as passwords, transaction addresses, etc.
%Without the proper process, the leakage of sensitive data may happen.
We further identify three patterns related to the blockchain wallet. %in this category.
First, since secret keys of a blockchain wallet are critical,
%\textbf{P16} immediately wipes the memory via \texttt{memwipe()} (Monero \#4268) or \texttt{memory\_cleanse()} (Bitcoin \#10308) after generating some secrets, as shown in Table~\ref{tab:p16exp}.
\textbf{P16} immediately wipes the memory via \texttt{memwipe()} (Monero \#4268) or \texttt{memory\_cleanse()} (Bitcoin \#10308) after generating some secrets.
Second, the addresses in a wallet are also sensitive and should be kept in testnet or memory.
%P17, on the other hand, deals with the privacy of addresses.
%As addresses stored in an unencrypted file may be a risk, it is much safer to store them in memory.
For example, Monero \#3315 in \textbf{P17} adds a \texttt{create\_address\_file} option for the address generating function \texttt{generate()} to create an address file only in the testnet environment.
Similarly, Bitcoin \#787 keeps the address table in memory and only writes to file when necessary. 
Third, a blockchain wallet requires users to always input passwords for critical operations.
For example, Monero \#4791 in \textbf{P18} performs such password checks via \texttt{ask\_password()}.

\textbf{Other blockchain-specific patterns.}
From P19 to P21, we summarize the last three kinds of blockchain-specific patterns.
Specifically, \textbf{P19} checks the validity of a Stellar-specific concept called \textit{Quorum}, which represents a set of nodes that are sufficient to reach an agreement in the Stellar network~\cite{lokhava2019stellar}.
%For example, Stellar \#2233 (see Table~\ref{tab:p19exp} in Appendix~\ref{sec:appendix1}) and \#2209 check the sanity of Quorum via \texttt{isQuorumSetSane()}.
For example, Stellar \#2233 and \#2209 check the sanity of Quorum via \texttt{isQuorumSetSane()}.
\textbf{P20} is a RPC-related pattern, which restricts the gas cap of RPC calls.
%If the requested gas exceeds the cap limit via \texttt{RPCGasCap()}, the caller should be warned (see Ethereum \#19401 in Appendix~\ref{sec:appendix1}).
If the requested gas exceeds the cap limit via \texttt{RPCGasCap()}, the caller should be warned (see Ethereum \#19401).
%This can help protect against DoS, as without the cap, an RPC call may use up the computing resource.
The last pattern, \textbf{P21}, asks a blockchain client to gracefully shutdown itself when there are unfinished block synchronization and processing.
This can be done by setting a global blockchain lock via \texttt{CRITICAL\_REGION\_LOCAL()},
%\lxnote{the type and description for P21 in Table 7 make is sound like a generic bug for a DB, not specific to blockchain; can revise them a bit so that it's more blockchain-specific? E.g., 'Blockchain corruption'?}
as shown in Monero \#706.

\subsection{Applying the Obtained Patterns for Vulnerability Detection}
\label{subsec:apply_pattern}

%\TODO{\red{======================START}}

While the focus of this paper is not vulnerability detection, we demonstrate the impact of obtained patterns by applying them to detect the same kinds of vulnerabilities in other blockchain projects. 
Specifically, in the blockchain world, it is normal for new blockchains to fork or partially reuse the code of classic blockchains, such as Bitcoin and Ethereum.
These ``forked'' blockchains thus could encounter similar vulnerabilities that appeared in Bitcoin and Ethereum.
Here we demonstrate a simple \textit{direct} search of the vulnerable clones and leave a \textit{variant} search to the future work. %based on control and data flows to our future work.

Among the top 100 cryptocurrencies\footnote{Based on the market cap at \url{https://coinmarketcap.com/} on 15 June 2021.}, we identified that 11 blockchains were forked from Bitcoin or had similar codebase as Bitcoin, including the rank \#6 Dogecoin, \#11 Bitcoin Cash, \#12 Litecoin.
For each blockchain $b$ and a given vulnerability pattern $p$, we leverage the vulnerable file $f$, vulnerable function method $m$, and surrounding code $c$ to first locate a clone of the original (unpatched) code (of $p$).
We then determine whether the cloned code is vulnerable or not by checking if a patch (of $p$) has been applied.
We use the pattern P3, P7, P11, P13, and P14 (see \mysec\ref{subsec:code_pattern}) that caused Bitcoin vulnerabilities in recent years to detect their cloned ones in the 11 projects.
The results are worrisome: six projects are affected by at least one vulnerability, and two projects, Dogecoin and Bitcoin SV, even suffer from all five kinds of vulnerabilities.
Among a total of the 20 vulnerabilities discovered, only ten use the same file and function name as the original vulnerability.
%TODO add more details?
This suggests the importance of generalized code patterns over the exact signatures relying on file names or function names.

We have reported all the 20 vulnerabilities to their corresponding vendors (via Email, GitHub, Discord, and Dash's Bounty) and offered them fix suggestions in late June and early July 2021.
%The bug report is avaliable at \url{https://tinyurl.com/fse227-bugreport}.
A summary of our vulnerability reporting is available at \url{https://tinyurl.com/fse-227}.
Dogecoin promptly confirmed all of our five reports and planned to fix them in their next minor release version.
Zcash, Horizen, and Ravencoin also confirmed our reports and are coordinating with their developers for fixes.
Dash checked our report on P3 and believed that they had applied a different patch by sorting inputs and outputs based on BIP69. 
Bitcoin SV replied to two of our reports on P7 and P14; however, they seem not keen to fix them.
We are still waiting for Bitcoin SV's replies on the other three reports.

\section{Related Work}
\label{sec:related}
% \vspace{-0.5ex}

%TODO In this section, we go through some related works on blockchain vulnerabilities, empirical vulnerability studies, and mining-based vulnerability detection.

\textbf{Blockchain vulnerability research.}
%Saad et al.~\cite{saad2019exploring} explored the details of past attacks on blockchains and the connections between these attacks.
%Homoliak et al.~\cite{homoliak2020security} provided a standardized four-layer model for studying blockchain vulnerabilities and categorized some blockchain incidents by applying this model.
%Besides these studies on summarizing blockchain attacks, 
Existing blockchain security studies mainly focus on smart contract vulnerability detection and transaction- or network-level analysis of the blockchains.
For \emph{smart contract vulnerability detection}, both static and dynamic program analysis tools have been proposed.
For instance, Oyente~\cite{luu2016smarter}, Zeus~\cite{Kalra2018ZEUSAS}, Securify~\cite{tsankov2018securify}, Gigahorse~\cite{grech2019gigahorse}, and ETHBMC~\cite{ETHBMC20} detected vulnerable smart contracts via symbolic execution, while ContractFuzzer~\cite{jiang2018contractfuzz} and ConFuzzius~\cite{ConFuzzius21} used fuzzing inputs to detect smart contract vulnerabilities, and Sereum~\cite{Sereum19} and SODA~\cite{SODA21} monitored run-time contract execution to detect on-chain attacks in modified EVMs.
%TODO Moreover, learning-based tools, such as SmartEmbed~\cite{SmartEmbed20}, ESCORT~\cite{ESCORT21}, and AMEVulDetector~\cite{AMEVulDetector21} were also recently invented.
For \emph{transaction-level analysis}, Karame et al.~\cite{karme2012fastbitcoin} analyzed the double-spending resilience of Bitcoin fast payments. % and showed that those attacks, without significant overhead, can succeed with relatively high probability.
%Chen et al.~\cite{chen2020ethereumgraph} addressed three security issues in Ethereum by analyzing graphs built based on Ethereum transactions.
%TODO Chen et at.~\cite{EthereumGraphAnalysis18} performed a systematic study of Ethereum transactions via graph analysis.
TxSpector~\cite{Zhang2020TXSPECTORUA} studied Ethereum transactions by replaying historical transactions and recording EVM bytecode-level traces. %, and analyzing data- and control-dependencies.
DeFiPoser~\cite{DeFiPoser21} proposed methods for discovering profit-generating transactions in DeFi protocols just in time.
For \emph{network-level analysis}, Apostolaki et al.~\cite{apostolaki2017hijackbitcoin} analyzed routing attacks by hijacking BGP prefixes and showed that such attacks could delay the propagation of blocks without being detected.
%Targeting on another type of attack, 
Gao et al.~\cite{gao2019minebitcoin} showed that by power adjusting and bribery racing, attackers could increase their mining rewards.

\textbf{Mining-based vulnerability detection.}
Code clone detection is a long-standing research topic in the software engineering area~\cite{SourcererCC2016, CanNeuralClone2021, LogStatement2021, FunctionalCodeDetect2021, Clcdsa2019, DNNBinary2022}.
%, and it has been used by the security community for mining vulnerable code fragments (e.g.,~\cite{redebug_2012, CodePropertyGraph14, vuddy_2017, xiao2020mvp}). 
Existing approaches are mainly based on detecting duplicated token subsequences or identifying exact or similar subtrees in abstract syntax tree (AST) representations. 
For \emph{token-based} approaches, CCFinder~\cite{ccfinder_2002}, CP-Miner~\cite{cpminer_2004}, and ReDeBug~\cite{redebug_2012} are the representative work. %TODO, all of which first split code into token sequences and then calculate the similarity of the code tokens for multilinguistic clone detection. 
Recently, a token-based approach, VUDDY~\cite{vuddy_2017}, generated code fingerprints via abstraction and normalization to speed up code clone detection. 
For \emph{tree-based} approaches, e.g., DECKARD~\cite{deckard_2007} and CloneDR~\cite{clonedr_1998}, they considered code’s structural information by generating ASTs and embedding them into a vector space for similarity comparison. 

\vspace{-1ex}
\section{Conclusion}
\label{sec:conclusion}
\vspace{-0.5ex}

In this paper, we conducted the first empirical study of blockchain system vulnerabilities and their security patches using four representative blockchains. %TODO, Bitcoin, Ethereum, Monero, and Stellar.
To enable this study, we proposed a vulnerability filtering framework to effectively identify 1,037 vulnerabilities and their 2,317 patches from 34,245 issues/PRs and 85,164 commits on GitHub.
Based on this unique dataset, we performed three levels of analyses. %TODO, namely file-level vulnerable module categorization, text-level vulnerability type clustering, and code-level vulnerability pattern analysis.
%In particular, we further applied the generated patterns to detect the same kinds of vulnerabilities in other top blockchains.
Our analysis revealed three key findings of blockchain system vulnerabilities, including
(i) the modules related to consensus, wallet, and networking are highly susceptible, each with over 200 issues;
(ii) around 70\% of blockchain vulnerabilities are in traditional types, but we also identify four new types specific to blockchains;
and (iii) we are able to obtain 21 blockchain-specific vulnerability patterns that check unique blockchain attributes and validate various blockchain statuses, and demonstrate that they can be applied to detect similar vulnerabilities in other top blockchains.
In the future, we will perform a thorough detection of blockchain system vulnerabilities based on the patterns extracted in this paper.

% \vspace{-0.5ex}
\section*{Acknowledgements}
\vspace{-0.5ex}
We thank all the reviewers of this paper for their detailed and constructive comments.
This work is partially supported by a direct grant (ref. no.
4055127) from The Chinese University of Hong Kong.

%{\small \bibliographystyle{acm}
%\bibliography{main}}
\balance
\bibliographystyle{ACM-Reference-Format}
\bibliography{main}

\end{document}